\documentclass[preprint,aps]{revtex4}
\usepackage{amsmath,amssymb,epsf} 
\begin{document}

\newtheorem{lemma}{Lemma}[section]
\newtheorem{pro}[lemma]{Proposition}
\newtheorem{cor}[lemma]{Corollary}
\newtheorem{theorem}[lemma]{Theorem}

\newcommand{\cierto}[1]{\fbox{#1}\label{#1}}

\newcommand\bu{\boldsymbol u}\newcommand\bU{\boldsymbol U}
\newcommand\ba{\boldsymbol a}\newcommand\bb{\boldsymbol b}
\newcommand\bn{\boldsymbol n}\newcommand\bk{\boldsymbol k}
\newcommand\bx{\boldsymbol x}\newcommand\by{\boldsymbol y}
\newcommand\bz{\boldsymbol z}
\newcommand\bv{\boldsymbol v}\newcommand\bV{\boldsymbol V}
\newcommand\bw{\boldsymbol w}\newcommand\bW{\boldsymbol W}
\newcommand\be{\boldsymbol e}\newcommand\br{\boldsymbol r}
\newcommand\bo{\boldsymbol 0}\newcommand\boo{\boldsymbol o}
\newcommand\bOO{\boldsymbol O}

\newcommand\btau{\boldsymbol \tau}
\newcommand\bh{\boldsymbol h}\newcommand\bg{\boldsymbol g}

\newcommand\bcalT{\boldsymbol{\cal T}}\newcommand\bcalL{\boldsymbol{\cal L}}
\newcommand\bPhi{\boldsymbol{\Phi}}\newcommand\bvarphi{\boldsymbol{\varphi}}
\newcommand\bOmega{\boldsymbol{\Omega}}

\newcommand\ip{\boldsymbol\cdot}\newcommand\dip{\boldsymbol :}
\providecommand\bnabla{\boldsymbol{\nabla}}

\newcommand\er{\mbox{\rm e}}\newcommand\ir{\mbox{\rm i}}



\title{A phenomenological model of weakly damped Faraday waves and the
  associated mean flow}

\author{Jos\'{e} M. Vega}
\affiliation{E.T.S.I. Aeron\'auticos, Universidad Polit\'ecnica de Madrid. 
Plaza Cardenal Cisneros, 3. 28040 Madrid. SPAIN.}
\author{Sten R\"udiger and Jorge Vi\~nals}
\affiliation{School of Computational Science and Information
Technology, Florida State University, Tallahassee, Florida 32306-4120,
USA.}

\date{\today}

\def\etal{\text{et al.}}
\def\tand{\text{\quad and\quad}}
\def\at{\text{\quad at\quad}}
\def\cc{\hbox{c.c.}}
\font\ochob=cmbx8
\font\ocho=cmr8

\begin{abstract}

A phenomenological model of parametric surface waves (Faraday waves) 
is introduced in the limit of small viscous dissipation that accounts
for the coupling between surface motion and slowly varying streaming
and large scale flows (mean flow). The primary bifurcation of the model is to a
set of standing waves (stripes, given the functional form of the model
nonlinearities chosen here). Our results for the secondary
instabilities of the primary wave show that
the mean flow leads to a weak destabilization of the base state
against Eckhaus and Transverse Amplitude 
Modulation instabilities, and introduces a new longitudinal 
oscillatory instability which is absent without the coupling. We compare our
results with recent one dimensional amplitude equations for this
system systematically derived from the governing hydrodynamic equations.

\end{abstract}

\maketitle

\def\tru#1{\cierto{#1}\raise2truemm\hbox{\sc{#1}}}
\def\rin{{\rm in}}
\def\at{{\rm at}}
\def\and{{\rm and}}
\def\for{{\rm for}}
\def\cc{{\rm c.c.}}

\def\HOT{{\rm HOT}}
\def\Rey{{\rm Re}}


\section{Introduction}

The purpose of this paper is to couple a phenomenological order
parameter model of parametric surface waves in the limit of weak
viscous dissipation to slowly varying flows (mean flows). To date,
most theories of parametric surface waves near onset have neglected
such flows despite the
observation that their effect is of the same order as other cubic
nonlinear conservative terms retained. The coupling to the
phenomenological model presented here allows
us to discuss the simplest consequences that these flows have in a laterally
unbounded geometry, namely shifts in thresholds of secondary
instabilities of the base pattern of standing waves, and the
appearance of a new longitudinal oscillatory instability.

When a layer of an incompressible fluid is vibrated periodically along
the direction normal to the free surface at rest, it can exhibit
parametrically driven surface waves, also known as Faraday waves
\cite{re:faraday31,re:miles90,re:cross93,re:gollub99}. Just above the
primary instability of the planar free surface, a set of standing
surface waves emerge leading to a stationary pattern with a symmetry
that depends on the physical
parameters of the fluid and the frequency of the forcing
\cite{re:kudrolli96,re:binks97,re:binks97b,re:westra03}. Intricate
phenomena appear in the limit of weak viscous dissipation in which
nonlinear wave interactions responsible for wave saturation and
pattern selection are dominated by triad resonant interactions  
\cite{re:edwards94,re:zhang94,re:zhang97}. Whereas the first
bifurcation away from planarity is to a set of standing waves in which
mean flow effects are absent, mean flows are expected to be
important in determining the stability of the primary waves, and more generally in
weakly damped systems. In this latter case, standing wave amplitude
equations can be expected to be valid only very close to onset.

Current weakly nonlinear theory is restricted to the small region above
threshold in which standing waves are stable, a state in which mean
flows identically vanish. However, the contribution from mean flows to
the equations governing
the slow evolution of the surface waves can be of the same order as the
standard cubic nonlinear and conservative terms which are always
retained. Thus weakly nonlinear corrections to surface waves and mean flows
must be considered simultaneously, which has not been done in three
dimensions in the
limit considered below (see \cite{re:lapuerta02} for the analysis of
this limit in two dimensions 
and  \cite{re:vega01,re:higuera01,re:higuera02}
for the analysis of related limits). The effects considered here are
not unlike other known phenomenology that includes the streaming flow
produced by fixed surface waves, 
\cite{re:liu77,re:leibovich81,re:craik85,re:iskandarani91} and
references therein, and the evolution of surface waves in the presence
of a fixed vortical flow \cite{re:milewsky95,re:mashayek98}.

A consistent introduction of mean flow effects into the amplitude
equations for Faraday waves requires explicit consideration of special limits
that involve the physical dimensions of the container. We specifically
focus here on the case of a fluid depth that is logarithmically large compared to
the wavelength, and derive  a set of evolution equations for the
surface waves and the associated  mean flow in the double limit of
small viscosity and large aspect ratio (the ratio between the lateral
size of the container and the wavelength). We find
two separate contributions to mean flow, namely an inviscid
contribution arising from the slowly varying motion of the free surface,
similar to the one appearing in classical Davey-Stewartson
models \cite{re:davey74}, and a viscous one resulting from a slowly
varying shear stress produced by the oscillatory boundary layer
attached to the free surface. This latter contribution describes
vorticity transport (by viscous diffusion or convection) from the
boundary layer into the bulk \cite{re:longuet53}.

An important simplification in our analysis is that the cubic nonlinear
terms of the phenomenological model are chosen so as to lead to a
stripe pattern above
onset instead of a square pattern as experimentally observed in
the limit of weak viscous dissipation. While it is a simple matter to modify
the functional form of the cubic term to produce square patterns,
we have chosen to first clarify the effect of mean flows on
slow modulations of a stripe pattern. There is no satisfactory theory
at present that can account for the interaction between slow spatial
modulation of the waves and mean flows in three dimensions, and the
case of stripes is
considerably simpler than other symmetries involving a larger number of
plane wave components at onset.

\section{Formulation}

We consider a fluid layer of unperturbed depth $d^*$ supported by a
horizontal plate that is vibrating vertically with an amplitude $a^*$ and a
frequency $2\omega^*$, where the superscript $^*$ 
denotes dimensional quantities. In order to facilitate
comparison with related results in \cite{re:zhang97,re:zhang98}, we use for
adimensionalization the characteristic time $\omega^{*-1}$ and 
length $k^{*-1}$, where the wavenumber $k^*$ is  
related to $\omega^*$ by the inviscid dispersion relation
\begin{equation}
\omega^{*2}=g^*k^{*}+\sigma^* k^{*3}/\rho^*,\label{b1}
\end{equation}
in terms of the gravitational acceleration $g^*$, the surface tension
$\sigma^*$ and the density $\rho^*$, all assumed constant.
Here we are assuming that the wavelength $k^{*-1}$ is (at least,
somewhat) small compared with 
the depth of the container. The
resulting dimensionless continuity and Navier-Stokes equations in
a reference frame attached to the vibrating container, with the $z=0$ plane at
the unperturbed free surface, are
\begin{alignat}{1}
&\bnabla\ip\bu+\partial_z
w=0,\label{b2}\\
&\partial_t\bu-w(\bnabla w-\partial_z\bu)-\bu^\perp\bnabla\ip\bu^\perp=-\bnabla
p+\gamma(\bnabla^2\bu+\partial^2_{zz}\bu)/2,\label{b3}\\
&\partial_tw+\bu\ip(\bnabla w-\partial_z\bu)=-\partial_z
p+\gamma(\bnabla^2w+\partial^2_{zz}w)/2,\label{b4}
\end{alignat}
in $-d<z<h(x,y,t)$, with boundary conditions resulting from no slip at the
supporting plate,
\begin{alignat}{1}
&\bu=\bo,\quad w=0\quad\text{at
}z=-d,\label{b5}
\end{alignat}
and kinematic compatibility and equilibrium of
tangential and normal stresses at the free surface,
\begin{alignat}{1}
&\partial_t h+\bu\ip\bnabla h=w,\label{b6}\\
&\partial_z\bu+\bnabla
w-(\bnabla\bu+\bnabla\bu^\top)\ip\bnabla h+  
[2\partial_zw-(\partial_z\bu+\bnabla w)\ip\bnabla h]\bnabla h=0,\label{b7}\\
&p-(|\bu|^2+w^2)/2-[4a\sin2t+1-\Gamma]h+\Gamma\bnabla\ip[\bnabla
h/(1+|\bnabla h|^2)^{1/2}]\label{b8}\\
&\hskip20truemm=\gamma [\partial_zw-(\partial_z\bu+\bnabla w)\ip\bnabla h
+\left(\bnabla h\ip(\bnabla\bu+\bnabla\bu^\top)/2\right)\ip\bnabla h]/
(1+|\bnabla h|^2)\nonumber
\end{alignat}
at $z=h$. Here  
\begin{equation}
\bu=(u,v,0)\label{b8a}
\end{equation}
 and  $w$ are the horizontal and vertical velocity components, 
\begin{equation}
\bnabla=(\partial_x,\partial_y,0)\label{b8b}
\end{equation}
denotes the horizontal gradient, the superscript $\perp$ over 
a horizontal vector denotes the result of rotating the vector
$90^\circ$ counterclockwise, namely
\begin{equation}
\bu^\perp=(-v,u,0),\label{b8c}
\end{equation} 
and the superscript  $\top$ over a tensor denotes the transpose;
$p$ (=pressure$+(|\bu|^2+w^2)/2+[1-\Gamma+4a\sin(2t)]z$) is a
conveniently modified pressure, and  
$h$ is the (vertical) free surface deflection.
For simplicity we do not consider lateral walls, but impose periodic boundary
conditions in two horizontal directions, namely
\begin{equation}
\begin{gathered}
(\bu,w,p)(x+L_1,y,z,t) = (\bu,w,p)(x,y+ L_2,z,t) = (\bu,w,p)(x,y,z,t),\\
f(x+L_1,y,t)= f(x,y+ L_2,t)=f(x,y,t).\end{gathered}\label{b9}
\end{equation}
 And for convenience we
also consider the vertically integrated continuity equation
\begin{equation}
\partial_t
h+\bnabla\ip\left(\int_{-d}^h\bu\,dz\right)=0,\label{b10}
\end{equation}
obtained upon integration of (\ref{b2}) in $-d<z<h$ and substitution of (\ref{b6}).

\subsection{Multiple scale analysis: oscillatory and mean flows}

We consider next a specific range of parameters in which it is
possible to simplify the problem by separating fluid motion into a 
\lq\lq fast'' oscillatory
component, and a \lq\lq slow'' mean flow. In particular, we consider
the system of surface waves near onset, and in the limits of a very
large lateral surface and weak viscous dissipation.
The problem depends on the following dimensionless parameters: the dimensionless 
viscosity $\gamma=2\nu^* k^{*2}/\omega^*$ (with $\nu^*$ the kinematic viscosity), 
the gravity-capillary contribution
$\Gamma=\sigma^* k^{*3}/(\rho^*\omega^{*2})$, the forcing amplitude
$a=a^*k^*$, the container depth $d=d^{*}k^{*}$ and the aspect ratios
$L_1$ and $ L_2$; note that, according to (\ref{b1}), $0\leq\Gamma\leq1$ and
the extreme cases $\Gamma=0$ and $1$ correspond to the purely gravitational and
purely capillary limits respectively. The approximation below requires that 
(a) the aspect ratio of the container be 
large, (b) the surface waves be weakly damped and
 (c) exhibit a small wavelength compared to the container's depth and (d) a
 small steepness, which in turn  require that
\begin{equation}
L\gg1,\quad
d\gg1,\quad\gamma\ll1,\quad|\bnabla
h|\ll1,\quad a\ll1,\label{b11}
\end{equation}
where $L\leq\min\{L_1,L_2\}$. The large spatial scale set by the
(large) aspect ratio introduces a slow horizontal scale over which both
spatial and temporal
modulations are expected to occur. As suggested by the 2-D case 
\cite{re:lapuerta02}, this scale is expected 
to be determined (in the equations for the oscillatory flow associated
with the surface waves) 
by the balance between cubic nonlinearity and either (i) transport
with the group velocity or (ii) dispersion. And $d$ must be not too
large, see below.
For the sake of clarity we assume that $d$ is logarithmically large
compared to the remaining small parameters 
(namely $\gamma$, $a$ and $L^{-1}$) and
{\it we shall treat $ d$ as a $O(1)$ parameter}. In fact, for simplicity 
we consider the distinguished limit
\begin{equation}
\gamma^2\ll\er^{-d}\ll L^{-1}\sim\gamma\sim a\sim\varepsilon^2,\label{b11a}
\end{equation}
where  $\varepsilon$ is a measure of the surface wave amplitude, 
see (\ref{b12}) below. The estimates 
$\gamma \sim a \sim \varepsilon^2$ result from imposing that 
linear damping, cubic nonlinearity and parametric forcing be of the same order. 
Therefore we are implicitly assuming that the coefficient of the cubic
term is $O(1)$, which excludes  {\it triad resonances} \cite{re:zhang97}. 
If these are present,  the cubic coefficient becomes $O(\gamma^{-1})$
and a different scaling applies. In order to concentrate on the effects
of mean flows, we exclude triad resonances in what follows.

Under these assumptions we shall (implicitly) use a multi-scale
analysis in both (horizontal) space and time. But in order to make a
not too technical presentation and  to avoid obscuring the main ideas with 
a too involved notation, we shall use only one time variable and one
space variable in each horizontal direction. The basic (fast) scales  
involve $O(1)$ increments of $x$, $y$ or $t$. A magnitude $\psi$ that
exhibits these fast scales at leading order is such that  
\begin{alignat}{1}
|\partial\psi/\partial x|\sim|\psi|,\quad |\partial\psi/\partial y|\sim|\psi|
\quad\text{or}\quad |\partial\psi/\partial t|\sim|\psi|,\label{b11b}
\end{alignat} 
and it will be said to {\it depend strongly} on the 
associated variable $x$, $y$ or $t$. If instead the magnitude only
changes over the slower scale at leading order, namely if 
\begin{alignat}{1}
|\partial\psi/\partial x|\ll|\psi|,\quad |\partial\psi/\partial y|\ll|\psi|
\quad\text{or}\quad |\partial\psi/\partial t|\ll|\psi|,\label{b11c}
\end{alignat}
then the magnitude  will be said to {\it depend weakly} on the 
associated variable, $x$, $y$ or $t$. To proceed, we 
decompose the flow variables and the free surface
deflection into oscillatory and time-averaged parts, associated with
the surface waves and the mean flow (denoted hereinafter with the
superscripts $o$ and $m$), respectively, as
 \begin{alignat}{1}
&(\bu,w,p,h)=\varepsilon(\bu^o,w^o,p^o,h^o)+\varepsilon^2(\bu^m,w^m,p^m,h^m),
\label{b12}
\end{alignat}
where (i) the (oscillatory) flow variables associated with the surface
waves are required to be such that
\begin{equation}
\langle\bu^o\rangle^{ts}=\bo,\quad\langle w^o\rangle^{ts}=\langle 
p^o\rangle^{ts}=\langle h^o\rangle^{ts}=0,
\label{b13}
\end{equation}
with $\langle\cdot\rangle^{ts}$ standing here and hereafter for the time average in
the basic  oscillating period
\begin{equation}
\langle\psi\rangle^{ts}=(2\pi)^{-1}\int_t^{t+2\pi}\psi\,dt;\label{b14}
\end{equation}
and (ii) the variables associated with the mean flow are required to
depend weakly on time; more precisely we assume that
\begin{equation}
|\partial_t\bu^m|\sim\varepsilon^2|\bu^m|,\quad|\partial_tw^m| \sim
\varepsilon^2| w^m|,
\quad|\partial_tp^m|\sim\varepsilon^2|p^m|,\quad
|\partial_th^m|\sim\varepsilon^2|h^m|,\label{b13a}
\end{equation}
where we are anticipating the timescale for the slow evolution of the
mean flow, $t\sim\varepsilon^{-2}$. Also we anticipate that the
re-scaled flow variables $\bu^o$, \ldots, $h^o$, $\bu^m$, \ldots,
$h^m$ are at most of order unity; see below. The
mean flow is described in terms of the time-averaged
velocity $\varepsilon^2\bu^m$, which is the {\it Eulerian} velocity
and does not coincide in general with the velocity associated with the
time average of the trajectories of fluid elements. The latter is the 
{\it Lagrangian} mean velocity, or {\it mass transport} velocity (denoted here as
$\varepsilon^2\bu^{mt}$) which is the appropriate velocity to analyze mean
trajectories of passive scalars (see e.g., \cite{re:umeki91,re:feng95}
in connection with chaotic advection \cite{re:aref02}). The difference
between the two is the {\it Stokes drift} so that, in the notation of
this paper, its horizontal and vertical components, scaled with
$\varepsilon^2$, are given by \cite{re:batchelor67}
\begin{equation}
\begin{gathered}
\bu^{mt}-\bu^m=\bu^{Sd} = \langle(\int^t\bnabla\phi\ip\bnabla)\bnabla\phi+
(\int^t\partial_z\phi)\partial_z\bnabla\phi\rangle^{ts},\\
w^{mt}-w^m=w^{Sd} = \langle(\int^t\bnabla\phi)\ip\bnabla
(\partial_z\phi)+(\int^t\partial_z\phi)\partial_{zz}\phi\rangle^{ts}
\end{gathered}\label{b14a}
\end{equation}
in a first approximation, where we are anticipating Eq. (\ref{b15})
below, and the operator $\int^t$ is defined as 
\begin{equation} 
\int^t\psi=\langle\int_{t_0}^t\psi\,dt\rangle^{to},\label{b14b}
\end{equation}
with $\langle\cdot\rangle^{to}$ standing hereinafter for {\it the 
time-oscillatory part}, defined as
\begin{equation} 
\langle\psi\rangle^{to}=\psi-\langle\psi\rangle^{ts}.\label{b14c}
\end{equation}
By definition, Eq. (\ref{b14b}) is independent of $t_0$.

\subsection{Oscillatory flow}

We begin by deriving the equations governing
the oscillatory flow associated with the surface waves which exhibits
a thin viscous
boundary layer of ${\cal O}(\sqrt{\gamma}$) thickness attached to the free
surface. In the bulk region outside this boundary layer the
oscillatory velocity components and the pressure are given by 
\begin{equation}
\begin{gathered}
\bu^o=\bnabla\phi+\varepsilon^2[(\int^t\partial_z\phi)(\bnabla w^m
-\partial_z\bu^m)+(\int^t\bnabla\phi)^\perp\bnabla\ip\bu^{m\perp}] +
O(\varepsilon^3), \\
\quad w^o=\partial_z\phi-\varepsilon^2(\int^t\bnabla\phi)\ip(\bnabla
w^m-\partial_z\bu^m)+O(\varepsilon^3),\quad p=-\partial_t\phi,
\end{gathered}\label{b15}
\end{equation}
as obtained upon substitution of Eq. (\ref{b12}) into
Eqs. (\ref{b3})-(\ref{b4}), where $\phi$ is the velocity potential. 
Thus the oscillatory flow is potential at leading
order, but not at order $\varepsilon^2$ which must be retained in what
follows. Substitution of Eq. (\ref{b15}) into Eq. (\ref{b2}) yields,
after some algebra,
\begin{alignat}{1}
\bnabla^2\phi+\partial^2_{zz}\phi
&+\varepsilon^2(\int^t\partial_z\phi)\bnabla\ip(\bnabla w^m
-\partial_z\bu^m)+\varepsilon^2\bnabla\ip[(\int^t\bnabla\phi)^\perp
\bnabla\ip\bu^{m\perp}]\nonumber\\
&-\varepsilon^2(\int^t\bnabla\phi)\ip\partial_z(\bnabla
w^m-\partial_z\bu^m)=0\quad\text{for }-d<z<0.\label{b16}
\end{alignat}
Here we are taking the upper boundary at the unperturbed free surface,
which can be done because $h$ is small. The boundary conditions at the
upper boundary must include  the effect of the vortical flow in the
boundary layer attached to the free surface on the oscillating flow in
the bulk. To the approximation relevant here this only requires to
replace the boundary condition (\ref{b7}) by 
(see, e.g., \cite{re:zhang97})
\begin{alignat}{1}
&\partial_t h+\bu\ip\bnabla h=w+W(x,y,t)\quad\text{at }z=h,\label{b16a}
\end{alignat}
where $W$ is given by 
\begin{alignat}{1}
&\partial_tW(x,y,t)=\gamma\bnabla^2(\partial_z\phi);\label{b16b}
\end{alignat}
this equation  can be integrated to obtain
\begin{equation} 
W=\gamma\int^t\bnabla^2(\partial_z\phi).\label{b16c}
\end{equation}
Thus, to the approximation relevant here Eq. (\ref{b10}) can be rewritten as
\begin{equation}
\partial_t
h+\bnabla\ip\left(\int_{-d}^h\bu\,dz\right)=\gamma\int^t\bnabla^2(\partial_z\phi),
\label{b16d}
\end{equation}
The boundary conditions for the oscillatory flow at the unperturbed
free surface are now obtained by a Taylor expansion from Eqs. 
(\ref{b8}), (\ref{b15}) and (\ref{b16d}), and are found to be
\allowdisplaybreaks{
\begin{alignat}{1}
&\partial_th^o-\partial_z\phi+\varepsilon^2(\int^t\bnabla\phi)\ip(\bnabla
w^m-\partial_z\bu^m)+\varepsilon\langle\bnabla\ip(h^o\bnabla\phi)\rangle^{to}
\nonumber\\
&\hskip25truemm+\varepsilon^2\bnabla\ip[h^m\bnabla\phi
+ h^o\bu^m+
(h^o)^2\partial_{z}\bnabla\phi/2]=\gamma\int^t\bnabla^2(\partial_z\phi),
\label{b17}\\
&\partial_t\phi+\varepsilon\langle h^o\partial^2_{tz}\phi\rangle^{to}
+\varepsilon^2
[h^m\partial^2_{tz}\phi-h^o\partial_zp^m+(h^o)^2\partial^3_{tzz} \phi/2]
+\varepsilon\langle|\bnabla\phi|^2+|\partial_z\phi|^2\rangle^{to}/2 \nonumber\\
&\hskip25truemm+\varepsilon^2[\bu^m\ip\bnabla\phi+h^o\partial_z(|\bnabla\phi|^2
+ |\partial_z\phi|^2 )/2]
+4a\langle h^o\sin2t\rangle^{to}\label{b18}\\
&\hskip40truemm+ (1-\Gamma)h^o -\Gamma\bnabla\ip[\bnabla
  h^o/(1+\varepsilon^2|
\bnabla h^o|^2)^{1/2}]
+\gamma\partial^2_{zz}\phi=0, \nonumber
\end{alignat}
where we are using Eq.(\ref{b40a}) below. 
The boundary condition (\ref{b5}) at the lower plate and the
periodicity conditions (\ref{b9}) yield
\begin{alignat}{1}
&\partial_z\phi=0\quad\text{at }z=-d,\label{b20}\\
&\begin{gathered}
\phi(x+L_1,y,z,t)= \phi(x,y+ L_2,z,t) = \phi(x,y,z,t),\\
h^o(x+L_1,y,t) =  h^o(x,y+ L_2,t) = h^o(x,y,t).\end{gathered} \label{b20A}
\end{alignat}

We are consistently neglecting terms of order $\varepsilon^3$ in Eqs. 
(\ref{b17})-(\ref{b18}) because of the approximations listed in Eq.
(\ref{b11a}), and taking into
account that those terms that are either (a) cubic in the oscillatory
flow variables or, (b) linear in both a oscillatory variable and a
slowly varying variable, exhibit zero temporal mean values at leading order.

Before proceeding any further we note that mean flow does not contribute to the 
averaged (in the time scale $t\sim1$) energy equation at leading
order, which is consistent with the fact that mean flow variables
(velocity and free surface deflection) are small compared to 
their counterparts in the oscillatory flow. The averaged energy
equation is obtained upon multiplication of Eq. (\ref{b16}) by 
$\partial_t\phi$, integration in $0<x<L_1,0<y<L_2,-d<z<0$, averaging 
over a period of oscillation, integration by parts repeatedly and  
substitution of Eqs. (\ref{b17})-(\ref{b18}) and (\ref{b20A}). We find
\begin{equation}
\frac{d\cal E}{dt}=-\int_0^{L_1}\int_0^{L_2}\gamma
\langle\partial_z(|\bnabla\phi|^2+|\partial_z\phi|^2)
+8a(\partial_th^o)\langle h^o\sin2t\rangle^{to}\rangle^{ts}\,dxdy
+{\cal O}(\gamma+a+\varepsilon^2),
\label{b20b}
\end{equation}
where $\cal E$ is the time averaged (kinetic plus potential)
energy. The first term in the right hand side of Eq. (\ref{b20b})
(except for a factor of two) is the classical result, first given by
Landau and Lifshitz \cite{re:landau76} that approximated viscous
dissipation by linear damping from the bulk potential flow (see also
\cite{re:zhang94,re:zhang97}). Note that mean flow variables (both
velocity and free surface deflection) are small compared to 
their counterparts in the oscillatory flow and do not contribute to
the energy at leading order. To obtain Eq. (\ref{b20b}) we  have taken
into account that $\bu^m$ and $w^m$ are
independent of $t$ at leading order, and that if $\psi$ and
$\varphi$ are $t$-periodic, of period $2\pi$
(as the variables associated with the oscillatory flow are to first
approximation), then to leading order we have
$\langle\varphi \int^t\psi\rangle^{ts}=-\langle\psi \int^t\varphi\rangle^{ts}
\quad\text{and}\quad\langle\psi \int^t\psi\rangle^{ts}=0$.

\subsection{Mean flow}

In order to obtain the equations and boundary conditions governing the
slowly varying flow we must take into account the oscillatory boundary
layer attached to the
free surface, which provides (at the edge of this layer) a slowly varying shear
stress that must be imposed as a boundary condition for the mean 
flow in the bulk. This forcing mechanism was first uncovered by
Longuet-Higgins \cite{re:longuet53}, who obtained an explicit expression
for the forcing shear stress produced by general boundary layers in 2-D. The
counterpart of this expression in 3-D (for a free surface of general shape) 
has been only obtained quite recently 
\cite{re:nicolas03}, although quasi-planar free surfaces (as the ones
considered here) were already considered in a not too well known work
\cite{re:liu77b}. With the notation of this paper, the general
formulae derived in \cite{re:nicolas03} yield
\begin{alignat}{1}
&\partial_z\bu^m+\bnabla
w^m=2\langle\bnabla\left(\bnabla\ip(h^o\bnabla\phi)\right)+(\bnabla
h^o\ip\bnabla)\bnabla\phi+(\bnabla^2\phi)\bnabla
h^o\rangle^{ts}\quad\text{at }z=0,
\label{b22}
\end{alignat}
where only the leading order contribution as $\gamma\to0$ and $\varepsilon\to0$ is
retained. The boundary layer attached to the free surface has no
effect on the other two boundary
conditions at the unperturbed free surface, which are directly  obtained
from Eqs. (\ref{b8}) and (\ref{b10}) to be
\begin{alignat}{1}
&p^m-(1-\Gamma)h^m+\Gamma\bnabla^2h^m=\langle h^o\partial^2_{tz}\phi + 
(|\bnabla\phi|^2
+|\partial_z\phi|^2)/2\rangle^{ts} \label{b23}\\
&\text{and}\quad\partial_th^m+\bnabla\ip(\int_{-d}^0\bu^m\,dz)=
-\bnabla\ip(\langle h^o\bnabla\phi\rangle^{ts})\quad\text{at }z=0,
\label{b24}
\end{alignat}
where we are only taking into account the leading order terms.
And from Eq. (\ref{b5}) we have no slip at the lower plate
\begin{alignat}{1}
&\bu^m=\bo,\quad w^m=0\quad\text{at
}z=-d,\label{b23a}
\end{alignat}
at leading order. We are neglecting the effect of the oscillatory 
boundary layer attached to the lower plate because its effect is quite
small (the horizontal component of the mean flow velocity near the
lower plate is proportional to the square of the vertical jump of the
horizontal component of the oscillatory velocity accross the lower
boundary layer \cite{re:batchelor67}, which is 
$O(\er^{-2d})\varepsilon^2\ll\varepsilon^2$ (Eq. (\ref{b11})); 
this in turn is small compared to the streaming flow 
velocity in the bulk, which is $O(\varepsilon^2)$).  
These boundary conditions show that mean flow
is forced by surface waves in two ways.  Those terms appearing in the 
right hand sides of Eqs. (\ref{b23}) and (\ref{b24}) provide an
inviscid forcing mechanism that by itself would provide an {\it
inviscid mean flow}, like that appearing in the Davey-Stewartson model 
\cite{re:davey74}. The right hand side of Eq. (\ref{b22}) instead
produces a forcing shear stress that drives a {\it viscous mean flow}, which 
is absent in the usual inviscid and nearly inviscid theories of
Faraday waves. Note that his forcing stress is generically non zero and 
independent of viscosity at leading order, fact that is well known but 
somewhat surprising because this effect is due to 
the oscillatory boundary layer, and is absent in the strictly inviscid
case. We remind the reader that the limit of vanishing 
viscosity is a {\it singular limit} which does commute with the limit 
$\varepsilon \to 0$. We could decompose the mean flow into its
inviscid and viscous parts, as is done in 
\cite{re:vega01,re:higuera01}, but for convenience this is not done here.
 
Finally, we substitute Eqs. (\ref{b12}) and (\ref{b15}) into
Eqs. (\ref{b2}) and (\ref{b3}), and take the time
average defined in Eq. (\ref{b14}) in the resulting
equations. Proceeding as we did to obtain Eq. (\ref{b20b}) 
we find, after some algebra
\begin{alignat}{1}
&\bnabla\ip\bu^m+\partial_z
w^m=0,\label{b25}\\
&\partial_t\bu^m-\varepsilon^2[(w^{Sd}+w^m)(\bnabla w^m-\partial_z\bu^m)
+(\bu^{Sd}+\bu^m)^\perp\bnabla\ip\bu^{m\perp}] \nonumber \\
&\hskip25truemm=-\bnabla
q^m+\gamma(\bnabla^2\bu^m+\partial^2_{zz}\bu^m)/2,\label{b26}\\
&\partial_tw^m+\varepsilon^2(\bu^{Sd}+\bu^m)\ip(\bnabla
w^m-\partial_z\bu^m)=
-\partial_z q^m+\gamma(\bnabla^2w^m+\partial^2_{zz}w^m)/2,\label{b27}
\end{alignat}
where $\bu^{Sd}$ and $w^{Sd}$ are the horizontal and vertical
components of the Stokes drift given by Eq. (\ref{b14a}), 
and the modified pressure $q^m$ is defined as
\begin{equation}
q^m=p^m+\varepsilon^2\langle\partial_z\phi(\int^t\bnabla\phi)\ip(\bnabla
w^m-\partial_z\bu^m)\rangle^{ts}.\label{b28}
\end{equation}
Finally the periodicity condition (\ref{b9}) yields
\begin{equation}
\begin{gathered}
(\bu^m,w^m,q^m)(x+L_1,y,z,t) = (\bu^m,w^m,q^m)(x,y+ L_2,z,t) = 
(\bu^m,w^m,q^m)(x,y,z,t),\\
h^m(x+L_1,y,t) = h^m(x,y+ L_2,t) = h^m(x,y,t).\end{gathered} 
\label{b29A}
\end{equation}

In order to estimate the magnitude of the various terms that depend on
the surface wave variables and force the mean flow, we assume that at
leading order the velocity potential $\phi$ and the free surface
deflection $h^o$ can be written as a superposition of plane waves 
\begin{equation}
\phi=\ir\er^z\sum_{n=-N}^{N}A_n\er^{{\scriptsize
    \ir}(t+\bk_n\ip\bx)}+\cc+
\ldots,\quad h^o=
\sum_{n=-N}^{N}A_n\er^{{\scriptsize \ir}(t+\bk_n\ip\bx)}+\cc+\ldots,\label{b30}
\end{equation}
where the complex amplitudes, 
$A_{-N}$,\ldots,$A_{N}$ are allowed to depend only on slow space and
time variables. The wavevectors $\bk_{-N}$,\ldots,$\bk_{N}$ correspond to
only $N$ directions because they are related in pairs as
\begin{equation}
\bk_{-n}=-\bk_n\quad\text{and}\quad|\bk_n|=1\quad\text{for} n=1,\ldots,N.
\label{b31}
\end{equation} 
Thus for each $n=1,\ldots,N$ the $n$-th and the $(-n)$-th waves 
counter-propagate along the same direction.
Note that each pair of counter-propagating waves builds a standing wave 
in the short timescale  if and only if $|A_n|=|A_{-n}|$, and 
the whole surface wave pattern is seen to be  standing in the short
time scale if and only if the 
following, more stringent condition holds 
\begin{equation}
A_{m}\bar A_n=\bar A_{-m} A_{-n}\quad\text{for all }m,n=1,\ldots,N.
\label{b32}
\end{equation}
Such  surface waves will be called {\it quasi-standing} below.

By using Eq. (\ref{b30}), the forcing terms in the boundary conditions 
Eqs. (\ref{b22})-(\ref{b24}) are written  as
\begin{alignat}{1} 
&\langle 2h^o\partial^2_{tz}\phi+|\bnabla\phi|^2+
|\partial_z\phi|^2\rangle^{ts} = \langle |\bnabla\phi|^2
-|\partial_z\phi|^2\rangle^{ts}=\sum_{m,n}(\bk_m\ip\bk_n-1)\bar A_m
 A_n\er^{{\scriptsize \ir}(\bk_n-\bk_m)\ip\bx}\nonumber\\
&\hskip20truemm +\sum_{m,n}\ir(A_n\bk_n\ip\bnabla\bar A_m
-\bar A_m\bk_m\ip\bnabla A_n)\er^{{\scriptsize \ir}(\bk_n-\bk_m)\ip\bx}+\cc
+O(L^{-2}), \label{b33}\\
&\bnabla\ip\langle h^o\bnabla\phi\rangle^{ts} = \langle\bnabla
\ip h^o\bnabla\phi\rangle^{ts}=-2\sum_{m}\bk_m\ip
\bnabla(|A_m|^2)+O(L^{-2}), \label{b34}\\
&\langle\bnabla\left(\bnabla\ip(h^o\bnabla\phi)\right)+(\bnabla
h^o\ip\bnabla)\bnabla\phi+(\bnabla^2\phi)\bnabla h^o\rangle^{ts}=\nonumber\\
&\hskip10truemm\sum_{m,n}(1+\bk_m\ip\bk_n)\bk_m\bar A_m A_n
\er^{{\scriptsize \ir}(\bk_n-\bk_m)\ip\bx}+\cc
+O(L^{-1}), \label{b35}
\end{alignat}
at $z=0$; and the Stokes drift 
velocity components in Eq. (\ref{b14a}) are written as
\begin{alignat}{1}
&\bu^{Sd}=\omega^{-1}\er^{2z} \sum_{m,n}(\bk_m\ip\bk_n+1)\bk_m\bar 
A_m A_n\er^{{\scriptsize \ir}(\bk_n -\bk_m)\ip\bx}+\cc+O(L^{-1}),
 \label{b36}\\
&w^{Sd}=2\omega^{-1}\er^{2z}\sum_{m}\bk_m\ip\bnabla(|A_m|^2)+O(L^{-2}), 
\label{b36a}
\end{alignat}
in $-d<z<0$.

Several remarks can be made about these boundary conditions. First,
the forcing term given in Eq. (\ref{b33}) depends only weakly on the horizontal
variables $x$ and $y$, namely
\begin{equation}
\langle \langle 2h^o\partial^2_{tz}\phi+|\bnabla\phi|^2
+|\partial_z\phi|^2 \rangle^{ts}\rangle^{hs}\sim L^{-1},\label{b37}
\end{equation}
where $\langle\cdot\rangle^{hs}$ is the horizontal average in
the short spatial scales
\begin{equation}
\langle\psi\rangle^{hs}=\left(\int_{\cal B}\,
dxdy\right)^{-1}\int_{\cal B}\psi \,dxdy.\label{b38}
\end{equation}
Here $\cal B$ is a ball of radius large compared to 1 but small
compared to $L$. Or, in terms of a horizontal Fourier transform 
with associated wavenumbers $\bk_{mn}\neq\bo$ if $(m,n)\neq(0,0)$, 
with $\bk_{00}=\bo$,
\begin{alignat}{1}
&\langle\psi\rangle^{hs}=\psi_{00}\quad\text{if}\quad
\psi=\sum_{m,n}\psi_{mn}\er^{{\scriptsize \ir}\bk_{mn}\ip\bx},\label{b39}
\end{alignat}
where $\psi_{mn}$ is allowed to depend weakly on $x$, $y$ and $t$  
(and strongly on $z$). 

Second, the forcing terms in Eqs. (\ref{b34}) and (\ref{b35}), and the Stokes
drift vanish at leading order if the surface wave pattern is  quasi-standing
\begin{alignat}{1}
&\langle\bnabla\left(\bnabla\ip(h^o\bnabla\phi)\right)+(\bnabla
h^o\ip\bnabla)\bnabla\phi+(\bnabla^2\phi)\bnabla h^o\rangle^{ts}\sim 
\bu^{Sd}\sim L^{-1},\nonumber\\
&\hskip10truemm\text{and} \quad\bnabla\ip\langle h^o\bnabla\phi\rangle^{ts}
\sim w^{Sd}\sim L^{-2}\quad\text{if Eq. (\ref{b32}) holds.} \label{b40}
\end{alignat}

Third, and according to Eqs. (\ref{b34}) and (\ref{b36a}), we have
\begin{equation}
\left|\left[\langle\bnabla\ip(h^o\bnabla\phi)\rangle^{ts}\right]_{z=0}\right|
=O(L^{-1}),\quad|w^{Sd}|=O(L^{-1}).\label{b48}
\end{equation}
By using the continuity equation (\ref{b25}), the boundary condition 
(\ref{b24}) can be rewritten as $\partial_th^m+w^m+\langle h^o
\bnabla\phi\rangle^{ts})=0$, which invoking Eqs. (\ref{b13a}) 
and (\ref{b48}) yields, at leading order,
\begin{equation}
w^m\sim L^{-1}\quad\text{at }z=0.\label{b40a}
\end{equation} 

\subsubsection{Short and long wave decomposition of the mean flow}

We next decompose the mean flow variables into a {\it  short wave} component
(or oscillatory in the horizontal direction) and a {\it long wave}
component (or slowly varying in the horizontal direction)
\begin{alignat}{1}
&(\bu^m,w^m,q^m,h^m)=(\bu^{mo},w^{mo},\varepsilon^2 q^{mo},h^{mo})
+(\bu^{ms},\varepsilon^2 w^{ms},q^{ms},h^{ms}),
\label{b41}
\end{alignat}
where the short wave component is such that
\begin{equation}
\langle\bu^{mo}\rangle^{hs}=\bo,\quad \langle w^{mo}\rangle^{hs}
=\langle q^{mo}\rangle^{hs}=\langle h^{mo}\rangle^{hs}=0.\label{b41a}
\end{equation}
The long wave components depend weakly on the horizontal variables
\begin{equation}
|\bnabla\bu^{ms}|\sim\varepsilon^2|\bu^{ms}|,\quad|\bnabla w^{ms}|
\sim\varepsilon^2|w^{ms}|,
\quad|\bnabla p^{ms}|\sim\varepsilon^2|p^{ms}|,
\quad|\bnabla h^{ms}|\sim\varepsilon^2|h^{ms}|,
\label{b41b}
\end{equation}
where we are assuming that the slow spatial scale for horizontal gradients of
the long wave mean flow is the same as that of the envelope of surface
waves, namely of the order of $L\sim\varepsilon^{-2}$ (see (\ref{b11a})).
Also, in Eq. (\ref{b41}) $\bu^{mo}\sim w^{mo}\sim q^{mo}\sim h^{mo}
\sim\bu^{ms}\sim w^{ms}\sim q^{ms}\sim h^{ms}\sim1$ and thus we are
anticipating the order of magnitude of all variables associated with
both mean flow components. 

The equations governing the short wave component of the mean flow are obtained 
by substituting Eq. (\ref{b41}) into (\ref{b22})-(\ref{b27}). The
short wave deflection $h^{mo}$ is given by
\begin{alignat}{1}
&-(1-\Gamma)h^{mo}+\Gamma\bnabla^2h^{mo}=
\left[\langle\langle h^o\partial^2_{tz}\phi+(|\bnabla\phi|^2
+|\partial_z\phi|^2)/2\rangle^{ts}\rangle^{ho}\right]_{z=0}, 
\label{b42}
\end{alignat}
with periodic boundary conditions
$ h^o(x+L_1,y,t) =  h^o(x,y+L_2,t) = h^o(x,y,t)$, and where
where $\langle\cdot\rangle^{ho}$ denotes the short wave component
\begin{equation}
\langle\psi\rangle^{ho}=\psi-\langle\psi\rangle^{hs}.
\label{b47b}
\end{equation}
Note that, according to Eq. (\ref{b37}), the horizontal mean value of
the right hand side of 
Eq. (\ref{b42}) vanishes at leading order, as required by volume conservation.

Short wave velocity and pressure $\bu^{mo}$, $w^{mo}$ and $q^{mo}$ are given by
\begin{alignat}{1}
&\bnabla\ip\bu^{mo}+\partial_z
w^{mo}=0,\label{b43}\\
&\partial_t\bu^{mo}-\varepsilon^2\langle w^{mo}(\bnabla w^{mo}-
\partial_z\bu^{mo}-\partial_z\bu^{ms})+(\bu^{Sd}+\bu^{mo}+\bu^{ms})^\perp
\bnabla\ip\bu^{mo\perp}\rangle^{ho}=\nonumber\\
&\hskip70truemm -\varepsilon^2 \bnabla
q^{mo}+\gamma(\bnabla^2\bu^{mo}+\partial^2_{zz}\bu^{mo})/2,\label{b44}\\
&\partial_tw^{mo}+\varepsilon^2\langle (\bu^{Sd}+\bu^{mo}+\bu^{ms})\ip
(\bnabla w^{mo} - \partial_z\bu^{mo}-\partial_z\bu^{ms})\rangle^{ho}= \nonumber\\
&\hskip70truemm-\varepsilon^2 \partial_z
q^{mo}+\gamma(\bnabla^2w^{mo}+\partial^2_{zz}w^{mo})/2,\label{b45}
\end{alignat}
in $-d<z<0$, with boundary conditions
\begin{alignat}{1}
&\bu^{mo}=\bo,\quad w^{mo}=0\quad\text{at }z=-d, \label{b46}\\
&\partial_z\bu^{mo}=2\langle \langle(\bnabla
h^o\ip\bnabla)\bnabla\phi+(\bnabla^2\phi)\bnabla h^o\rangle^{ts}\rangle^{ho},
\quad w^{mo}=0 \quad\text{at }z=0,\label{b47}\\
&\begin{gathered}
(\bu^{mo},w^{mo},q^{mo})(x+L_1,y,z,t) = (\bu^{mo},w^{mo},q^{mo})(x,y+ L_2,z,t) \\
= (\bu^{mo},w^{mo},q^{mo})(x,y,z,t),\\
h^{mo}(x+L_1,y,t) = h^{mo}(x,y+ L_2,t) = h^{mo}(x,y,t).
\end{gathered} \label{b47a}
\end{alignat}
where we have taken into account Eq. (\ref{b48}). Note that this
problem is decoupled from that giving $h^{mo}$ (Eq. (\ref{b42})).

In order to determine the equations governing the long wave component of the
mean flow, we first take into account that, according to the continuity
equation and the boundary condition (\ref{b23a}), the rescaled
vertical velocity component is given by
\begin{equation}
w^{ms}=-\varepsilon^{-2}\int_{-d}^z\bnabla\ip\bu^{ms}\,dz\label{b50}
\end{equation}
and, according to Eq. (\ref{b41b}), is of order unity as assumed above.
The problem giving $(\bu^{ms},q^{ms},h^{ms})$ becomes decoupled
from  $w^{ms}$  as we show now. From the momentum equation (\ref{b27}) we obtain
\begin{equation}
q^{ms}=q^{ms}(x,y,t),\label{b51}
\end{equation}
and then, by invoking Eq. (\ref{b48}), the momentum equation (\ref{b26}) yields
\begin{equation}
\partial_t\bu^{ms}-\varepsilon^2
\langle w^{mo}(\bnabla w^{mo}-
\partial_z\bu^{mo})+(\bu^{Sd}+\bu^{mo})^\perp
\bnabla\ip\bu^{mo\perp}\rangle^{hs} =
- \bnabla q^{ms}+\gamma\partial_{zz}\bu^{ms}/2\label{b54}
\end{equation}
in $-d<z<0$, with boundary conditions
\begin{alignat}{1}
&\bu^{ms}=\bo\quad\text{at }z=-d,\quad \partial_z\bu^{ms}
=2\langle \langle(\bnabla h^o\ip\bnabla)\bnabla\phi+
(\bnabla^2\phi)\bnabla h^o\rangle^{ts}\rangle^{hs},\label{b55}\\
&q^{ms}-(1-\Gamma)h^{ms}=0,\quad\partial_th^{ms}
+\bnabla\ip(\int_{-d}^0\bu^{ms}\,dz)=-\bnabla\ip(\langle\langle h^o
\bnabla\phi\rangle^{ts}\rangle^{hs}),\label{b56}
\end{alignat}
at $z=0$, resulting from Eqs. (\ref{b22})-(\ref{b23a}) and (\ref{b28}).
The periodic boundary conditions (\ref{b9}) lead to
\begin{equation}
\begin{gathered}
(\bu^{ms},w^{ms},q^{ms})(x+L_1,y,z,t) = (\bu^{ms},w^{ms},q^{ms})(x,y+ L_2,z,t)\\
= (\bu^{ms},w^{ms},q^{ms})(x,y,z,t),\\
h^{ms}(x+L_1,y,t) = h^{ms}(x,y+ L_2,t) = h^{ms}(x,y,t).
\end{gathered} 
\label{b56a}
\end{equation}

All terms appearing in Eqs. (\ref{b54})-(\ref{b56}) are of the same order because
$\gamma\sim\varepsilon^2$ (see Eq. (\ref{b11a})) and $\bu^{ms}$, $q^{ms}$ and
$h^{ms}$ satisfy Eqs. (\ref{b13a}) and (\ref{b41b}). The forcing term in
Eq. (\ref{b56}b) $\bnabla\ip\langle\langle h^o\bnabla\phi\rangle^{ts}\rangle^{hs}$,
is inviscid, and is the only forcing term that appears in the standard 
Davey-Stewartson model \cite{re:davey74} which involves a potential mean flow. 
This model is only valid for stripes patterns as described below
in Eqs. (\ref{c18})-(\ref{c22}), where this term is precisely the only
forcing term remaining in the right hand side of Eq. (\ref{c22}). On
the other hand, the forcing terms in the right hand sides of Eqs. (\ref{b47}) and 
(\ref{b55}) are necessarily viscous and drive a vortical flow.

Substituting Eqs. (\ref{b41}) into Eqs. (\ref{b16})-(\ref{b20}), we
obtain the following equation (after some algebra):
\begin{alignat}{1}
\bnabla^2\phi+\partial^2_{zz}\phi
&+\varepsilon^2(\int^t\partial_z\phi)\bnabla\ip(\bnabla w^{mo}
-\partial_z\bu^{mo})+\varepsilon^2\bnabla\ip[(\int^t\bnabla\phi)^\perp
\bnabla\ip\bu^{mo\perp}]\nonumber\\
&-\varepsilon^2(\int^t\bnabla\phi)\ip\partial_z(\bnabla
w^{mo}-\partial_z\bu^{mo}- \partial_{z}\bu^{ms})=0\label{b57}
\end{alignat}
for $-d<z<0$, and boundary conditions 
\begin{alignat}{1}
&\partial_th^o-\partial_z\phi+\varepsilon^2(\int^t\bnabla\phi)\ip(\bnabla
w^{mo}-\partial_z\bu^{mo}-\partial_z\bu^{ms})+\varepsilon\langle\bnabla\ip(h^o
\bnabla\phi)\rangle^{to}\nonumber\\
&\hskip10truemm+\varepsilon^2\bnabla\ip[(h^{mo}+h^{ms})\bnabla\phi
+ h^o(\bu^{mo}+\bu^{ms})+
(h^o)^2\partial_{z}\bnabla\phi/2]\nonumber\\
& \hskip 10truemm =\gamma\int^t\bnabla^2
(\partial_z\phi),
\label{b58}\\
&\partial_t\phi+\varepsilon\langle h^o\partial^2_{tz}\phi\rangle^{to}
+\varepsilon^2
[(h^{mo}+h^{ms})\partial^2_{tz}\phi-h^o\partial_zq^{ms}+(h^o)^2
\partial^3_{tzz}\phi/2] \nonumber\\
& \hskip 10truemm +\varepsilon\langle|\bnabla\phi|^2 + 
|\partial_z\phi|^2\rangle^{to}/2 + \varepsilon^2[(\bu^{mo}+\bu^{ms})\ip\bnabla\phi
+h^o\partial_z(|\bnabla\phi|^2+|\partial_z\phi|^2 )/2]
\nonumber\\
& \hskip 10truemm +4a\langle h^o\sin2t\rangle^{to}
+ (1-\Gamma)h^o -\Gamma\bnabla\ip[\bnabla
  h^o/(1+\varepsilon^2|
\bnabla h^o|^2)^{1/2}]
+\gamma\partial^2_{zz}\phi=0,
\label{b59}\\
&\partial_z\phi=0\quad\text{at }z=-d.
\label{b60}
\end{alignat}

This is the central result of this section. In the limit considered
(Eq. (\ref{b11a})), flow variables have been decomposed
into oscillatory and slowly varying parts (Eq.  (\ref{b12})),
where the oscillatory components are given by Eq. (\ref{b15}), with $\phi$
and $h^o$ given by Eqs. (\ref{b16})-(\ref{b20A}). 
The slowly varying component has been further decomposed into short
wave and long wave components in Eq. (\ref{b41}), with the short
wave component given by Eq. (\ref{b42}), 
and the long wave given  by Eqs. (\ref{b54})-(\ref{b56}). 
Thus the coupled evolution of surface waves and mean flow is given by 
Eqs. (\ref{b42}), (\ref{b43})-(\ref{b47a}),
(\ref{b54})-(\ref{b56a}), and  (\ref{b57})-(\ref{b60}). 
This system of equations still includes the full
3-D Navier-Stokes equations (\ref{b44}) and (\ref{b45})  with a large Reynolds
number based on the horizontal size (although a $O(1)$ Reynolds number
based on the containers depth).

The numerical solution of this coupled problem remains quite
complicated, and further simplifications are necessary to make it
tractable. We discuss in the following section a hierarchy of
simplified models that are based on various physical
assumptions and in some cases on ad hoc approximations. 

\section{Approximations to the coupled mean flow-surface wave
  equations}
\label{sec:approximations}

\subsection{Stripes}
\label{sec:stripes}

In this particular case the equations derived above simplify substantially. 
Unfortunately in the nearly-inviscid limit we are considering 
stripes are not the selected pattern in 3D  \cite{re:chen99},   
except in the purely gravity wave limit of $\Gamma\ll 1$. This would 
require that the basic 
wavelength $2\pi/k^{*}$ be large compared to the capillary length 
$\ell_c=\sqrt{\sigma/(\rho g_0)}$, which is of the order of $3$ mm for
water. We consider first the 
strict gravitational limit of $\Gamma=0$.
If only one stripe is present at each point then we can take 
$N=1$ in Eqs. (\ref{b34})-(\ref{b36a}) at this point, and 
 the short wave  parts of $\bu^{Sd}$ and $\langle(\bnabla
h^o\ip\bnabla)\bnabla\phi+(\bnabla^2\phi)\bnabla h^o\rangle^{ts}$ vanish in
Eqs. (\ref{b43})-(\ref{b47}). This implies that
the  short wave part of the mean flow is unforced, and thus admits the solution
$\bu^{mo} = \bo,\; w^{mo} = 0,\; q^{mo}=\text{constant}$
which we assume is globally stable. Then for large times such that the
short wave component vanishes, the remaining variables are given by 
Eqs. (\ref{b16}), (\ref{b17})-(\ref{b20}), (\ref{b42})
and  (\ref{b54})-(\ref{b56}). By also invoking Eq. (\ref{b41b}), 
these latter equations can be rewritten as
\begin{alignat}{1}
&\bnabla^2\phi+\partial^2_{zz}\phi
+\varepsilon^2(\int^t\bnabla\phi)\ip\partial_{zz}\bu^{ms}=0,\label{c15}\\
&\partial_t\bu^{ms} =-\bnabla h^{ms}+\gamma\partial_{zz}\bu^{ms}/2\label{c16}
\end{alignat}
in $-d<z<0$, with boundary conditions
\begin{alignat}{1}
&\partial_z\phi=\bu^{ms}=\bo,\label{c17}
\end{alignat}
at $z=-d$ and 
\begin{alignat}{1}
&\partial_th^o-\partial_z\phi-\varepsilon^2(\int^t\bnabla\phi)\ip
\partial_z\bu^{ms}+\varepsilon\langle\bnabla\ip(h^o\bnabla\phi)\rangle^{to}
\nonumber\\
&\hskip20truemm+\varepsilon^2\bnabla\ip[(h^{mo}+h^{ms})\bnabla\phi
+ h^o\bu^{ms}+ (h^o)^2\partial_{z}\bnabla\phi/2]=\gamma\int^t\bnabla^2
(\partial_z\phi),
\label{c18}\\
&\partial_t\phi+\varepsilon\langle h^o\partial^2_{tz}\phi\rangle^{to}
+\varepsilon^2
[(h^{mo}+h^{ms})\partial^2_{tz}\phi-\partial_zq^{ms}+(h^o)^2
\partial^3_{tzz}\phi/2] \nonumber\\
& \hskip 20truemm
+\varepsilon\langle|\bnabla\phi|^2+|\partial_z\phi|^2
\rangle^{to}/2 + \varepsilon^2[\bu^{ms}\ip\bnabla\phi
+h^o\partial_z(|\bnabla\phi|^2+|\partial_z\phi|^2 )/2]
\nonumber\\
& \hskip 20truemm +4a\langle h^o\sin2t\rangle^{to}+ h^o 
+\gamma\partial^2_{zz}\phi=0,\label{c19}\\
& \partial_z\bu^{ms}
=2\langle \langle(\bnabla h^o\ip\bnabla)\bnabla\phi+
(\bnabla^2\phi)\bnabla h^o\rangle^{ts}\rangle^{hs},\label{c21}\\
&\partial_th^{ms}
+\bnabla\ip(\int_{-d}^0\bu^{ms}\,dz)=
-\bnabla\ip(\langle\langle h^o\bnabla\phi\rangle^{ts}\rangle^{hs}),
\label{c22}
\end{alignat}
at $z=0$. The periodic boundary conditions 
in the horizontal variables resulting from Eq. (\ref{b9}) are 
\begin{equation}
\begin{gathered}
(\phi,h^{o})(x+L_1,y,z,t)=(\phi,h^{o})(x,y+ L_2,z,t)=(\phi,h^o)(x,y,z,t),\\
(h^o,h^{mo},h^{ms})(x+L_1,y,z,t)=(h^o,h^{mo},h^{ms})(x,y+
L_2,z,t)=(h^o,h^{mo},h^{ms})(x,y,z,t).\end{gathered}\label{c23}
\end{equation}
These equations are considerably simpler as they only include
the heat equation, Eq. (\ref{c16}), instead of the
full continuity and Navier-Stokes equations, but yet allow
significant variation in stripe orientation.  

\subsection{Linear approximation for the mean flow}

The mean flow equations  and boundary
conditions Eqs. (\ref{b43})-(\ref{b47}) and (\ref{b54})-(\ref{b56}) are linear in
the mean flow variables except for convective terms. If these are
neglected then the mean flow equations are rewritten as Eq. (\ref{b42}) and 
\begin{alignat}{1}
&\bnabla^2\phi+\partial^2_{zz}\phi
+\varepsilon^2(\int^t\partial_z\phi)\bnabla\ip(\bnabla w^{mo}
-\partial_z\bu^{mo})+\varepsilon^2\bnabla\ip[(\int^t\bnabla\phi)^\perp
\bnabla\ip\bu^{mo\perp}]\nonumber\\
&\hskip30truemm-\varepsilon^2(\int^t\bnabla\phi)\ip\partial_z(\bnabla
w^{mo}-\partial_z\bu^{mo}-\partial_{z}\bu^{ms})=0,\label{d3}\\
&\bnabla\ip\bu^{mo}+\partial_z
w^{mo}=0,\label{d4}\\
&\partial_t\bu^{mo}= -\varepsilon^2 \bnabla
q^{mo}+\gamma(\bnabla^2\bu^{mo}+\partial^2_{zz}\bu^{mo})/2,\label{d5}\\
&\partial_tw^{mo}=-\varepsilon^2 \partial_z
q^{mo}+\gamma(\bnabla^2w^{mo}+\partial^2_{zz}w^{mo})/2,\label{d6}\\
&\partial_t\bu^{ms} =-\bnabla q^{ms}+\gamma\partial_{zz}\bu^{ms}/2
\quad\text{in }-d<z<0,\label{d6a}
\end{alignat}
in $-d<z<0$, with boundary conditions
\begin{equation}
\partial_z\phi=0,\quad \bu^{mo}=\bu^{ms}=\bo,\quad w^{mo}=w^{ms}=0,\label{d7}
\end{equation}
at $z=-d$, and 
\begin{alignat}{1}
&\partial_th^o-\partial_z\phi+\varepsilon^2(\int^t\bnabla\phi)\ip(\bnabla
w^{mo}-\partial_z\bu^{mo}-\partial_z\bu^{ms})+\varepsilon\langle\bnabla\ip(h^o
\bnabla\phi)\rangle^{to}
\nonumber\\
&\hskip20truemm+\varepsilon^2\bnabla\ip[(h^{mo}+h^{ms})\bnabla\phi
+ h^o(\bu^{mo}+\bu^{ms})+
(h^o)^2\partial_{z}\bnabla\phi/2]
\nonumber\\
& \hskip 20truemm =\gamma\int^t
\bnabla^2(\partial_z\phi),
\label{d7a}\\
&\partial_t\phi+\varepsilon\langle h^o\partial^2_{tz}\phi\rangle^{to}
+\varepsilon^2
[(h^{mo}+h^{ms})\partial^2_{tz}\phi-h^o\partial_zq^{ms}+(h^o)^2
\partial^3_{tzz}\phi/2]
\nonumber\\
& \hskip 20truemm +\varepsilon\langle|\bnabla\phi|^2+
|\partial_z\phi|^2\rangle^{to}/2
+\varepsilon^2[(\bu^{mo}+\bu^{ms})\ip\bnabla\phi
\nonumber\\
& \hskip 20truemm +h^o\partial_z(|\bnabla\phi|^2+|\partial_z\phi|^2 )/2]
+4a\langle h^o\sin2t\rangle^{to}\nonumber\\
&\hskip20truemm+ (1-\Gamma)h^o -\Gamma\bnabla\ip[\bnabla
  h^o/(1+\varepsilon^2|
\bnabla h^o|^2)^{1/2}]
+\gamma\partial^2_{zz}\phi=0,
\label{d7b}\\
& w^{mo}=0,\quad\partial_z\bu^{mo}=2\langle\langle(\bnabla
h^o\ip\bnabla)\bnabla\phi+(\bnabla^2\phi)\bnabla
h^o\rangle^{ts}\rangle^{ho},
\label{d8}\\
&\partial_z\bu^{ms}
=2\langle \langle(\bnabla h^o\ip\bnabla)\bnabla\phi+
(\bnabla^2\phi)\bnabla h^o\rangle^{ts}\rangle^{hs},
\label{d10}\\
&q^{ms}-(1-\Gamma)h^{ms}=0,\quad\partial_th^{ms}
+\bnabla\ip(\int_{-d}^0\bu^{ms}\,dz)=-\bnabla\ip(\langle\langle
h^o\bnabla\phi
\rangle^{ts}\rangle^{hs}),\label{d11}\\
&-(1-\Gamma)h^{mo}+\Gamma\bnabla^2h^{mo}=
\langle\langle h^o\partial^2_{tz}\phi+(|\bnabla\phi|^2
+|\partial_z\phi|^2)/2\rangle^{ts}\rangle^{ho}, \label{d11a}
\end{alignat}
at $z=0$; and periodic boundary conditions in the horizontal direction
as in Eqs.(\ref{b47a}) and (\ref{b56a}).

This linear approximation exactly provides the first bifurcated branch 
of standing waves (SWs) from the planar base state with associated
mean flow that is unforced 
(see Eq. (\ref{b40})) and thus identically vanishes at large times. 
This approximation is also exact for the linear stability of the SWs and, 
in particular, for the instability threshold of this branch, namely
the threshold amplitude (if it is finite) for the appearance of
transverse amplitude modulations (TAMs) \cite{re:kudrolli96}. Furthermore,
we expect that the neglected convective terms do not play a
significant qualitative role in subsequent 
bifurcated branches of TAMs, at least near threshold. In addition 
this approximation is almost exact for stripes because convective
terms can be neglected in this case, as already explained 
in Sec. \ref{sec:stripes}.

\subsection{Two dimensional approximation}

We introduce next a (drastic) single mode approximation for the $z$
dependence of the mean flow variables in Eqs. (\ref{d3})-(\ref{d11}). 
We write horizontal velocities as
\begin{equation}
\bu^{mo}=g(z)\bU^{mo},\quad
q^{mo}=g(z)Q^{mo},\quad\bu^{ms}=g(z)\bU^{ms}, \quad q^{mo}=g(z)Q^{ms}, 
\label{d20}
\end{equation}
where the function $g$ is such that
\begin{alignat}{1}
&g(-d)=g'(0)=0,\quad \int_{-d}^0g(z)^2\,dz=1,\quad g(0)>0,\quad 
\int_{-d}^0g(z)\,dz>0. \label{d21}
\end{alignat}
This function is otherwise arbitrary and can be selected to yield the
best approximation to the vertical velocity profiles. A reasonable choice is
\begin{alignat}{1}
g(z)=\sqrt{2/d}\sin[\pi(z+d)/(2d)],
\label{d22}
\end{alignat}
which satisfies
\begin{equation}
g''=-\lambda g, \quad\text{with } \lambda= \frac{\pi^{2}}{4 d^{2}}
\label{d23}
\end{equation}
By using this simplification Eqs. (\ref{d4})-(\ref{d5}), (\ref{d6a})
and (\ref{d11}) reduce to
\begin{alignat}{1}
&\bnabla\ip\bU^{mo}=0,\label{d24}\\
&\partial_t\bU^{mo}= -\varepsilon^2 \bnabla
Q^{mo}+\gamma(\bnabla^2\bU^{mo}-\lambda\bU^{mo})/2 \nonumber \\
&\hskip25mm+\beta_1\gamma[\langle\langle(\bnabla
h^o\ip\bnabla)\bnabla\phi+(\bnabla^2\phi)\bnabla h^o
\rangle^{ts}\rangle^{ho}]_{z=0} ,\label{d25}\\
&\partial_t\bU^{ms} =-\bnabla Q^{ms}-\gamma\lambda\bU^{ms}/2
+\beta_1\gamma
[\langle\langle(\bnabla
h^o\ip\bnabla)\bnabla\phi+(\bnabla^2\phi)\bnabla
h^o\rangle^{ts}\rangle^{hs}]_{z=0},
\label{d26}\\
&\beta_1 Q^{ms}-(1-\Gamma)h^{ms}=0,\quad\partial_th^{ms}
+\beta_2\bnabla\ip\bU^{ms}
+\bnabla\ip(\langle\langle
h^o\bnabla\phi\rangle^{ts}\rangle^{hs})_{z=0}=0,
\label{d27}
\end{alignat}
with
\begin{alignat}{1}
&\beta_1=g(0)>0,\quad \beta_2=\int_{-d}^0 g(z)\,dz>0.
\label{d27a}
\end{alignat}
Note that if $g$ is given by Eq. (\ref{d22}), then 
\begin{alignat}{1}
&\beta_1=\sqrt{2/d},\quad \beta_2=\sqrt{8 d}/\pi.
\label{d27b}
\end{alignat}
The complete set of equations also includes Eq. (\ref{b42})
and the following equations and boundary conditions which follow
from Eqs. (\ref{d3}), (\ref{d7a}) and  (\ref{d7b})
\begin{alignat}{1}
&\bnabla^2\phi+\partial^2_{zz}\phi
+\varepsilon^2g''(z)(\int^t\bnabla\phi)\ip(\bU^{mo}+\bU^{ms})=0,
\label{d29}
\end{alignat}
in $-d<z<0$, with $\partial_z\phi=0$ at $z=-d$, and
\begin{alignat}{1}
&\partial_th^o-\partial_z\phi
+\varepsilon\langle\bnabla\ip(h^o\bnabla\phi)\rangle^{to}+\varepsilon^2
\bnabla\ip[(h^{mo}+h^{ms})\bnabla\phi
+\beta_1 h^o(\bU^{mo}+\bU^{ms})]
\nonumber\\
&\hskip 10truemm +\varepsilon^2\bnabla\ip[ (h^o)^2\partial_{z}\bnabla\phi/2]
=\gamma\int^t\bnabla^2(\partial_z\phi),
\label{d30}\\
&\partial_t\phi+\varepsilon\langle h^o\partial^2_{tz}\phi\rangle^{to}
+\varepsilon^2 [(h^{mo}+h^{ms})\partial^2_{tz}\phi+(h^o)^2\partial^3_{tzz}\phi/2]
\nonumber\\
& \hskip 10truemm
+\varepsilon\langle|\bnabla\phi|^2+|\partial_z\phi|^2
\rangle^{to}/2 +\varepsilon^2[\beta_1(\bU^{mo}+\bU^{ms})\ip\bnabla\phi
+h^o\partial_z(|\bnabla\phi|^2+|\partial_z\phi|^2 )/2] 
\nonumber\\
& \hskip 10truemm +4a\langle h^o\sin2t\rangle^{to}
+ (1-\Gamma)h^o -\Gamma\bnabla\ip[\bnabla
  h^o/(1+\varepsilon^2|
\bnabla h^o|^2)^{1/2}]
+\gamma\partial^2_{zz}\phi=0,\label{d31}
\end{alignat}
at $z=0$. Equations (\ref{b42}), and (\ref{d24})-(\ref{d31}) must be
integrated with periodic boundary conditions in the horizontal
directions, as above. 
Note that the mean flow does not contribute to the averaged energy 
equation at leading order (see Eq. (\ref{b20b})).
 
\subsection{A phenomenological description}
\label{sec:phen_model}

We finally discuss the simplest possible approximation to this problem
by considering a phenomenological model of Faraday waves that
qualitatively describes its primary bifurcation and secondary
instabilities \cite{re:zhang95}. It involves a complex order parameter
$\psi$ that satisfies
\begin{equation}
\partial_{t} \psi = - \gamma \psi + \ir f \bar\psi + 3 \ir \left(
1 + \nabla^{2} \right) \psi /4+ (\ir-\gamma\alpha) |\psi|^{2} \psi.
\label{d40}
\end{equation}
A derivation of this equation is given in the Appendix. The order
parameter $\psi$ is a linear combination of the free surface deflection 
and (a vertical average of) the velocity potential, 
\begin{equation}
h^o = \psi\er^{-{\scriptsize \ir}\omega t} +\cc ,\quad  \phi= -\ir
\psi \er^{-{\scriptsize \ir} \omega t}+ \cc,
\quad\text{or}\quad\psi=\er^{-{\scriptsize \ir} \omega
  t}(h^o+\ir\phi)/2.
\label{d41}
\end{equation} 
Despite its simplicity, the order parameter model qualitatively describes 
some of the features of the Faraday instability: The linear dispersion 
relation coincides with that of the fluid in the limit of low viscous damping, 
and the model exhibits a primary bifurcation to a standing wave  
solution near threshold, which can be either 
sub- or supercritical at threshold depending on the wavenumber. Also,
stationary solutions are in turn destabilized against amplitude and 
phase modulation instabilities. For sufficiently high supercriticalities, the 
solutions of Eq. (\ref{d40}) exhibit spatio temporal chaos.
Consistent with the weakly dissipative limit we are considering in
this paper, we must assume that
\begin{equation}
\gamma\ll1,\quad f\ll1,\quad|\alpha|\sim1.
\label{d42}
\end{equation}

This model has been already used by Kiyashko \etal \;
\cite{re:kiyashko96} to understand how mean flow effects might induce rotating 
patterns in a Faraday wave experiment. Conjecturing that rotation was
somehow due to the mean flow produced by surface waves, they added a
convective term $-\bu\ip\bnabla\psi$ to the right hand side 
of Eq. (\ref{d40}), where $\bu$ was a velocity field  that was given 
independently of $\psi$; thus ignoring any coupling between surface
waves and mean flow.
 
Here we shall add a similar term to the right hand side of
Eq. (\ref{d40}) but, given the analysis above, $\bu$ evolves with 
the surface waves according to a phenomenological equation with
the appropriate symmetries. First we replace Eq. (\ref{d40}) with
\begin{equation}
\partial_{t} \psi = - \gamma \psi + \ir f \bar\psi + 3 \ir \left(
1 + \nabla^{2} \right) \psi /4+ (\ir-\gamma\alpha) |\psi|^{2} \psi
-\beta_1(\bU^{mo}+\bU^{ms})\ip\bnabla\psi,\label{d43}
\end{equation}
as suggested by Eqs. (\ref{d30})-(\ref{d31}), (\ref{d40}) and
(\ref{d41}). Note that we are not including any dependence on $h^{mo}$
and $h^{ms}$, because this is beyond the scope of this
phenomenological model. The coupling term {\it is not conservative}
for general initial conditions, which is not optimal. However, this term 
may be seen to lead to a conservative contribution at leading order
for solutions that are linear combinations of plane waves (as in Eq. 
(\ref{b30}) above). Since $\psi$ is intended here to only 
model the spatially oscillatory part of the flow, we require that 
\begin{equation}
\langle\psi\rangle^{to}=0.
\label{d43a}
\end{equation}
The contribution from the mean flow appears through $\bU^{mo}$ and
$\bU^{ms}$, which are the short and long wave 
components of the mean flow defined above. Their evolution is given by 
Eqs. (\ref{d24})-(\ref{d27}) but replacing $h^o$ 
and $\phi$ by $\psi$ according to Eqs. (\ref{d40})-(\ref{d41})
\begin{alignat}{1}
&\bnabla\ip\bU^{mo}=0,\label{d44}\\
&\partial_t\bU^{mo}= -\varepsilon^2 \bnabla
Q^{mo}+\gamma(\bnabla^2\bU^{mo}-\lambda\bU^{mo})/2+
\beta_1\gamma\langle\ir(\bnabla
\psi\ip\bnabla)\bnabla\bar\psi+\ir(\bnabla^2\bar\psi)\bnabla \psi+\cc
\rangle^{ho},
\label{d45}\\
&\partial_t\bU^{ms} =-\bnabla
Q^{ms}-\gamma\lambda\bU^{ms}/2+\beta_1\gamma
\langle\ir(\bnabla
\psi\ip\bnabla)\bnabla\bar\psi+\ir(\bnabla^2\bar\psi)\bnabla \psi+\cc
\rangle^{hs},
\label{d46}\\
&\beta_2 Q^{ms}-(1-\Gamma)h^{ms}=0,\quad\partial_th^{ms}
+\beta_2\bnabla\ip\bU^{ms}
+\bnabla\ip(\ir\langle \psi\bnabla\bar\psi \rangle^{hs}+c.c.)=0.\label{d47}
\end{alignat}
We also require that ({\it cf.} Eqs. (\ref{d23}) and (\ref{d27a})) 
\begin{equation}
\beta_1^4/\lambda=16/\pi^{2},\quad\beta_1\beta_2=4/\pi.
\label{beta}
\end{equation}
Equations (\ref{d45})-(\ref{d47}) imply that ${\bf U}^\text{mo}={\bf
U}^\text{ms}=0$ as  $\beta_1\to0$. Note also that the mean flow is
unforced if the surface waves are standing (the phase of $\psi$ is
independent of position) as all forcing terms in
Eqs. (\ref{d45})-(\ref{d47}) vanish. 


\section{Secondary instabilities of the phenomenological model}
\label{sec:sec_ins}

In order to obtain a qualitative picture of the effects of mean flows
on surface waves, we study in this section secondary instabilities 
of the base periodic solution of the order parameter model defined
by the coupled Eqs. (\ref{d40}) and (\ref{d44})-(\ref{d47}) of
Sec. \ref{sec:phen_model}. 
Generally speaking, we find that mean flows couple weakly to transverse 
phase modulations and hence do not appreciably modify the zig-zag boundary. 
Transverse amplitude modulations are affected by mean flows, the
latter generally being destabilizing. Mean flows also increase the region 
of instability against longitudinal perturbations (Eckhaus) and, more 
importantly, introduce a finite wave number longitudinal instability 
which for certain values of the parameters can render much of the 
parameter space in which periodic solutions exist unstable. This 
instability branch is of oscillatory nature, and arises at the merging point 
between the branch that corresponds to long wavelength longitudinal modes of 
$\psi$ and a hydrodynamic branch which is weakly damped as $k \rightarrow 0$.
 
The trivial solution $\psi = \bU^{mo} = \bU^{ms} = Q^{mo} = Q^{ms} = h^{ms} 
= 0$ becomes linearly unstable against a periodic perturbation of $\psi$ of 
wave number $q$ for $\mu > \mu_c(q) = \sqrt{1+[3(1-q^2)/4 \gamma]^2}
-1$, the neutral stability curve. $\mu$ is the control parameter defined 
as $\mu=(f-\gamma)/ \gamma$. 
The critical mode $q=1$ becomes unstable at $\mu=0$.

For small $\mu > 0$, stationary and periodic solutions exist that can be 
approximated by a single Fourier mode $\psi_q(x) = \alpha_q \mbox{ cos }( q x) 
\exp{(\ir \Theta_q)}$ with 
\begin{equation}
\alpha_q^2=\frac{q^2-1-\frac{4}{3} \alpha \gamma^{2} \pm\frac{1}{3} 
\sqrt{16 f^2(1+\alpha^{2}\gamma^{2})
-(4 \gamma+3 \alpha\gamma (q^2-1))^2}}{1+(\alpha \gamma)^{2}},
\label{eq:psi_stat}
\end{equation}
where the $\pm$ sign stands for ${\rm sign}(1-q^2+4\alpha\gamma^{2}/3)$, 
and $\Theta_{q}$ satisfies $\mbox{sin } 2 \Theta_q=(1 + 3 \alpha 
\alpha_{q}^{2}/4)\gamma/f$,  
$\mbox{cos } 2 \Theta_q=\frac{3}{4}(q^2-1-\alpha_q^2)/f$. 
Note that the  bifurcation at threshold is  
subcritical if $q^2>1+3\alpha \gamma^{2}/4\simeq1$ (recall that $\gamma$ is small) 
and supercritical otherwise.
This solution for the order parameter leads to 
vanishing driving terms in the mean flow Eqs. (\ref{d44})-(\ref{d47}); hence 
all mean flow variables remain zero for the base, periodic solution.  

In order to address the stability of the stationary solution 
(\ref{eq:psi_stat}) against longitudinal perturbations, we introduce
\begin{eqnarray}
\psi&=&A_0[\exp{(\ir q x)}+\exp{(-\ir q x)}+a^{++}\exp{(\ir (q+k) x)}+
a^{+-} \exp{(\ir (q-k) x)} \nonumber \\
& &  +a^{-+} \exp{(\ir (k-q) x)}+a^{--}\exp{(-\ir (q+k) x)}]
\label{eq:psi_pert}
\end{eqnarray}
(where $A_0=\frac{1}{2} \alpha_q \exp{(\ir \Theta_q)}$),
together with the corresponding perturbations of the mean flow variables
\begin{equation}
U^{ms}_x=u^+ \exp{(\ir k x)}+ c.c.,
\label{eq:ums_pert}
\end{equation}
\begin{equation}
h^{ms}=c^+ \exp{(\ir k x)}+ c.c.,
\label{eq:h_pert}
\end{equation}
and
\begin{equation}
Q^{ms}=d^+ \exp{(\ir k x)}+ c.c.
\label{eq:q_pert}
\end{equation}
where $\bU^{mo}=0$ as seen from the incompressibility condition
(\ref{d44}), and the definition of  $\bu^{mo}=0$ (\ref{b41a}), which
require that $U^{mo}_x=0$ and $\langle \bU^{mo}\rangle^{hs}=0$, respectively.

By inserting Eq. (\ref{eq:psi_pert}) into the nonlinear terms 
of Eqs. (\ref{d46}) and (\ref{d47}), 
\begin{alignat}{1}
{\bf N}^{v}=&\ir(\bnabla
\psi\ip\bnabla)\bnabla\bar\psi+\ir(\bnabla^2\bar\psi)\bnabla \psi+\cc, \\
{\bf N}^{i}=&\ir\psi\bnabla\bar\psi +\cc,
\end{alignat}
and retaining only the long wavelength components we obtain
\begin{alignat}{1}
\langle N^{v}_{x}
\rangle^{hs} = &\langle 2 \ir \partial_x \psi \partial_{x}^{2} 
\bar{\psi}+\cc\rangle^{hs} \label{d49} \\
=&2 |A_0|^2 [(q+a^{++}(q+k) \er^{{\scriptsize \ir} k x}+a^{+-}(q-k) 
\er^{-{\scriptsize \ir} k x})
(q^2+\bar{a}^{++}(q+k)^2 \er^{-{\scriptsize \ir} k x} \nonumber \\
&+\bar{a}^{+-}(q-k)^2 \er^{{\scriptsize \ir} k x})
+(-q+a^{-+}(k-q) \er^{{\scriptsize \ir} k x}-a^{--}(q+k) 
\er^{-{\scriptsize \ir} k x}) \nonumber \\
&(q^2+\bar{a}^{-+}(k-q)^2 \er^{-{\scriptsize \ir} k x}+
\bar{a}^{--}(q+k)^2 \er^{{\scriptsize \ir} k x})]+\cc, \nonumber\\
\langle N^{i}_{x}\rangle^{hs} =&
\langle \ir \psi \partial_x \bar{\psi}
+\cc\rangle^{hs} \label{d52} \\
=&|A_0|^2 [(1+a^{++} \er^{{\scriptsize \ir} k x}+a^{+-}
  \er^{-{\scriptsize \ir} k x})
(q+\bar{a}^{++}(q+k) \er^{-{\scriptsize \ir} k x}+\bar{a}^{+-}(q-k) 
\er^{{\scriptsize \ir} k x})\nonumber \\
&+(1+a^{-+} \er^{{\scriptsize \ir} k x}+a^{--} \er^{-{\scriptsize \ir} k x})
(-q+\bar{a}^{-+}(k-q) \er^{-{\scriptsize \ir} k x}-\bar{a}^{--}(q+k) 
\er^{{\scriptsize \ir} k x})]+\cc,
\nonumber
\end{alignat}
respectively, where $N^{(v,i)}_{x}$ denote the $x$-components of the
nonlinear terms ${\bf N}^{(v,i)}$.
Equations for the perturbation amplitudes $u^{+}$, $c^+$, 
and $d^+$ are derived by linearizing Eqs.~(\ref{d46}), (\ref{d47}) with 
respect to all the perturbation amplitudes, after substitution of
Eqs. (\ref{d49}) and (\ref{d52}), and by extracting those terms 
that are proportional to $\er^{{\scriptsize \ir} k x}$:
\begin{alignat}{1}
\partial_t u^+&=-\ir k d^+ -\frac{\gamma \lambda}{2} u^++2 \beta_1 \gamma |A_0|^2
[(q+k)(2 q^2+q k)(a^{++}-\bar{a}^{--}) \nonumber \\
&\hskip10mm+(q-k)(2 q^2-q k)(\bar{a}^{+-}-a^{-+})], 
\label{d55} \\
\beta_2 d^+&=(1-\Gamma) c^+, \label{d56} \\ 
\partial_t c^+&=-\beta_2 \ir k u^+- \ir k |A_0|^2
[(2q+k) (a^{++}-\bar{a}^{--})+(2q-k)(\bar{a}^{+-}-a^{-+})]. 
\label{d57} 
\end{alignat}
Finally, the governing equations for the perturbation amplitudes of the 
order parameter are obtained by linearizing Eq. (\ref{d43}) and extracting 
the amplitudes of the Fourier modes $\exp{\ir (\pm q \pm k) x}$. For example, 
the governing equation for $a^{++}$ is
\begin{eqnarray}
\partial_t a^{++}&=&-\gamma a^{++}+\ir f \frac{\bar{A}_0}{A_0} 
\bar{a}^{--}+\frac{3 \ir}{4} 
(1-(q+k)^2) a^{++}\\
&&+(\ir-\alpha \gamma) |A_0|^2(4 a^{++}+2\bar{a}^{--}+
\bar{a}^{+-}+2 a^{-+}) - \beta_1 \ir q u^+. \nonumber
\label{d58}
\end{eqnarray}
Similar equations result for the other three amplitudes.

We now have a system of six first order ordinary differential
equations which is linear in the perturbation
amplitudes $a^{\pm \pm}$, $u^+$, and $c^+$. The matrix of right hand side 
coefficients is denoted by $A(q,k,\epsilon,\dots)$, and 
is a function of the wavenumbers of the 
base solution $q$ and of the perturbation $k$, of the control
parameter $\epsilon$, 
and of the other parameters of the model. The base solution becomes unstable
when the real part of any eigenvalue of this matrix becomes positive.
We have numerically obtained the eigenvalues of the matrix A, and determined 
the region of stability of the base solution. Two types of instabilities 
are possible: a standard long wavelength Eckhaus instability which depends on the 
mean flow, and a finite wavenumber oscillatory instability, which is
completely due to the mean flow. As was the case
in the asymptotically exact equations for one dimensional
Faraday waves \cite{re:lapuerta02}, this latter instability only
occurs with nonzero mean flow. Both instabilities will be further discussed below.

The stability of periodic solutions against
transverse amplitude and phase perturbations can be studied in a
similar fashion. Given that
$\partial_y \psi_q = 0$ in the base state with zero mean flow,
terms involving the $y$ components of the mean flow will be of second
order in the amplitudes of the perturbation and hence
only the components $U_x^{mo}$ and $U_x^{ms}$ need to be perturbed.
However, and in contrast to the case of a longitudinal perturbation,
both short and long wavelength components of the mean flow need to be
included. Furthermore, the equations for $U_x^{mo}$ and $U_x^{ms}$ 
decouple at linear order, and they can be analyzed separately.
We show next that only the $U_x^{mo}$ part modifies the transverse
amplitude modulation (TAM) instability line
whereas the zig-zag line is not affected by either component.

We start by considering the short wavelength component of the 
mean flow velocity $U_x^{mo}$, and introduce the following perturbation for 
the order parameter,
\begin{eqnarray}
\psi & = & A_0 \left[ \exp{(\ir q x)}+\exp{(-\ir q x)} +a^{++}
\exp{(\ir (q x+k y))}+ a^{+-}\exp{(\ir (q x-k y))} + \right. \nonumber \\
& & \left. + a^{-+}\exp{(\ir (-qx+ky))}+a^{--}\exp{(-\ir (qx+ky))} \right] ,
\label{eq:psi_pert2}
\end{eqnarray} 
\begin{equation}
U^{mo}_x=v^{++} \exp{(\ir (2 q x+k y))}+ v^{+-} \exp{(\ir (2 q x-k y))}
+\cc,
\end{equation}
and
\begin{equation}
Q^{mo}=p^{++} \exp{(\ir (2 q x+k y))}+ p^{+-} \exp{(\ir (2 q x-k y))}
+\cc
\end{equation}
The amplitude of the mode $\er^{{\scriptsize \ir} 2 q x}$ in the
$x$-component of the nonlinear forcing term 
of Eq. (\ref{d45}) is
\begin{alignat}{1}
\langle N^{v}_x \rangle^{ho}
& =\langle\it[\partial_x \psi \partial_{x}^{2} \bar{\psi}+
\partial_y\psi\partial_{xy}\bar{\psi}
+(\partial_{x}^{2}\bar{\psi}+\partial_{y}^{2}\bar{\psi})\partial_x\psi]+\cc
\rangle^{ho} 
\label{d59}\\
&= |A_0|^2 q k^2[(\bar{a}^{-+}-a^{+-}) \er^{-{\scriptsize \ir} k y}+
(\bar{a}^{--}-a^{++}) \er^{{\scriptsize \ir} k y}]
\er^{{\scriptsize \ir} 2 q x}+\cc,
\label{eq:nvx2}
\end{alignat}
where in Eq. (\ref{eq:nvx2}) only terms linear in $a^{\pm\pm}$ and
$\bar{a}^{\pm\pm}$ have been kept.
The pressure $Q^{mo}$ is calculated by taking the divergence of  Eq. (\ref{d45}),
\begin{equation}
0=-\nabla^2 Q^{mo} +\beta_1 \gamma (\partial_x N^{v}_x +\partial_y N^{v}_y),
\end{equation}
and is thus eliminated by using 
\begin{equation}
0=(4 q^2+k^2) p^{++} +\beta_1 \gamma \ir 2 q |A_0|^2 q k^2 (\bar{a}^{--}-a^{++}),
\end{equation}
and a similar equation for $p^{+-}$
(here we have used the fact that one obtains $\langle N^{v}_y 
\rangle^{ho}=0$ at linear order).
We then derive the following linear system of equations for the
perturbation amplitudes,
\begin{alignat}{1}
\partial_t a^{++}=&-\gamma a^{++}+\ir f \frac{\bar{A}_0}{A_0} 
\bar{a}^{--}+\frac{3 \ir}{4} 
(1-q^2-k^2) a^{++}\nonumber \\
&+(\ir-\alpha\gamma) 
|A_0|^2(4 a^{++}+2\bar{a}^{--}+\bar{a}^{+-}+2 a^{-+})
+ \beta_1 \ir q v^{++}. 
\label{d60} \\
\partial_t v^{++}=&-\frac{\gamma}{2} (4 q^2+k^2+\lambda) v^{++}+
\beta_1 \gamma |A_0|^2  \frac{q k^4}{4 q^2+k^2}
(\bar{a}^{--}-a^{++}), 
\label{d61}
\end{alignat}
with similar equations for $a^{+-}$, $a^{-+}$, $a^{++}$, and $v^{+-}$.

A transverse amplitude modulation (TAM) is defined by the 
linear combinations $b_1=a^{++}+a^{+-}+a^{-+}+a^{--}$ and 
$v_1=\Im(v^{++}+v^{+-})$. From Eqs. (\ref{d60}) and (\ref{d61})
we find a closed system
\begin{eqnarray}
\partial_t b_1&=&-\gamma b_1+\ir f \frac{\bar{A}_0}{A_0} \bar{b}_1 +
\frac{3 \ir}{4}(1-q^2-k^2) b_1
+\frac{3}{4} \alpha^2_q(\ir-\alpha \gamma)(2 b_1+\bar{b}_1)-2 \beta_1 q v_1, 
\label{d62}\\
\partial_t v_1&=&-\frac{\gamma}{2}(4 q^2+k^2+\lambda) v_1-\frac{\beta_1 
\gamma}{4} \alpha_q^2 \frac{q k^4}{4 q^2+k^2} \Im \,b_1. 
\label{d63}
\end{eqnarray}
The relevance of mean flows on this perturbation is demonstrated in
Fig. \ref{comp} showing the stability boundary
in the plane $(\beta_{1},\mu)$ at fixed wavenumber $q$ and
model parameters. With increasing mean flow coupling $\beta_{1}$,
the region of stability of the base solution (region above the dashed line
in the figure) decreases. 

Alternatively, a transverse phase modulation (zig-zag) is given 
by $b_2=a^{++}-a^{+-}-a^{-+}+a^{--}$ and $v_2=\Im(v^{++}-v^{+-})$.
We find in this case,
\begin{eqnarray}
\partial_t b_2&=&-\gamma b_2+\ir f \frac{\bar{A}_0}{A_0} \bar{b}_2 + 
\frac{3 \ir}{4}(1-q^2-k^2) b_2
+\frac{1}{4} \alpha^2_q(\ir-\alpha \gamma)(2 b_2+\bar{b}_2)-2 \beta_1 q v_2, 
\label{d64}\\
\partial_t v_2&=&-\frac{\gamma}{2}(4 q^2+k^2+\lambda) v_2-\frac{\beta_1 
\gamma}{4} \alpha_q^2  \frac{q k^4}{4 q^2+k^2} \Im \, b_2. 
\label{d65}
\end{eqnarray}
From the eigenvalue equation of this system, it can be shown that
modifications to the eigenvalues due to $v_2$ appear at higher
order in $k$. Since the zig-zag instability occurs in the limit
$k \rightarrow 0$ we find the effect of mean flow to be negligible in
this case.

A similar analysis has been performed for the long wavelength component of 
the mean flow with perturbations of the form $U_x^{ms}=u^+\exp{(\ir k y)} + 
\cc$, $h^{ms}=c^+\exp{(\ir k y)}+ \cc$, and $Q^{ms}=d^+\exp{(\ir k y)} + 
\cc$. The order parameter is again given by Eq. (\ref{eq:psi_pert2}).
As shown in Fig. \ref{comp}, the contribution of the long wavelength 
component of the mean flow to transverse modulations is negligible.

We now turn to a summary of our numerical results about the base
periodic solution. Figure \ref{balloon} shows the
various stability boundaries for the special cases of $\beta_{1}=0$ 
(no mean flow) and $\beta_{1} = 0.5$. 
Other values of the parameters used are $\gamma=0.1$, $\alpha=0.5$,
and $\Gamma=0.8$. The values of $\beta_2$ and $\lambda$ depend on $\beta_1$ 
according to Eq. (\ref{beta}). Except for $\alpha$, these parameter values
correspond approximately to the values for the low-viscosity
experiments of Kudrolli
and Gollub described in \cite{re:kudrolli96}. For instance, typical
experimental values of $\beta_{1} = \sqrt{2/d}$ (where $d$ is the
dimensionless height of the layer) are between 0.5 and 1.2.
Figure \ref{balloon} includes the neutral stability curve of the basic periodic 
solution and, since the primary bifurcation is subcritical for $q>1$, we 
have included the saddle node curve where the periodic solution bifurcates
(a subcritical bifurcation for $q>1$ as $\gamma \rightarrow 0$ has also
been found in a direct numerical solution of the governing fluid equations
in two dimensions \cite{re:chen00b}). 
The case $\beta_{1} = 0$ is shown as a reference, and it agrees
with the results of \cite{re:zhang95}. 

The range of base solutions that is stable against all perturbations 
considered here (Eckhaus, TAM, and zig-zag) is a small region close to 
threshold at $\mu=0$ between the TAM and zig-zag lines. 
Periodic solutions are stable against transverse perturbations below the 
dashed-dotted line in the figure (zig-zag, denoted \lq\lq Z"), and above the 
dashed line (TAM). Eckhaus perturbations have negative growth rate below the 
dotted line. We observe that with increasing $\beta_1$ both Eckhaus
and TAM curves are shifted so that larger regions in the 
($\mu$, $q$) space become destabilized with respect to TAM 
or Eckhaus perturbations. As discussed above, the zig-zag line is not 
affected by the mean flow.

We finally discuss a new oscillatory instability against longitudinal 
perturbations which is absent for $\beta_{1}=0$. The oscillatory nature of 
the instability is demonstrated in Fig. \ref{ew} that shows the real and 
imaginary parts of the corresponding critical eigenvalue branch as a function 
of the wavenumber of the perturbation $k$ for fixed $q, \mu$, and 
other model parameters. The imaginary part of the
eigenvalue $\sigma$ is not zero at the point in which $\Re (\sigma) = 0$.
This figure also shows that the instability occurs at small but finite 
wavenumber $k$, fact that has been confirmed by calculating the wave number
with largest growth rate both slightly above and below the instability
threshold at $\mu=0.1155$.

The inset in Fig. \ref{ew} illustrates the origin of the 
oscillatory instability. In the limit of $k \rightarrow 0$ there are
two distinct eigenvalue branches that have a small and negative real
part. The upper branch is marginal at $k = 0$ and is related to the
translational symmetry broken by the base state $\psi_{q}(x)$. On the other
hand, the mean flow velocity vanishes in the base state, originating
the lower branch which is weakly damped at $k = 0$. The damping rate
of the relevant mode for longitudinal perturbations ($U_x^{ms}$)
is $\gamma \lambda/2$ as can be seen from Eq. (\ref{d46}).
As $k$ increases, the two (real) eigenvalue branches merge
leading to a complex conjugated pair and to an oscillatory instability.

If the eigenvalue problem for the perturbations of Eckhaus type is expressed by
the real and imaginary parts of the coefficients $a^{\pm \pm}, u^+$, and $c^+$
two pairs of complex conjugated 
eigenvalues cross the imaginary axis at the instability point. Therefore, the 
effective dimensionality
of the critical subspace is four. This can be understood from the symmetry 
group of the system of Eqs.~(\ref{d55})-(\ref{d58}). These equations are 
invariant under a spatial reflection (exchanging $a^{++}$ with $a^{--}$, 
$a^{+-}$ with $a^{-+}$, and $u^+$ with $-\bar{u}^+$) and
rotation (acting on $a^{\pm +}$ and $u^+$ as
multiplication by $\exp{(\ir \Theta)}$ and on $a^{\pm -}$ by 
$\exp{(-\ir \Theta)}$, where $\Theta$ is arbitrary) that derive 
from the symmetry of the original equations. The corresponding 
symmetry group O(2) has two-dimensional irreducible representations
which in turn requires each eigenvalue to occur twice. 
At threshold, the eigenspace of the linear system is spanned by four
linearly independent eigenvectors, two for each pair of the
complex conjugate eigenvalues. Symmetric bifurcation theory shows that 
for generic bifurcations with
O(2) symmetry the nonlinear solution branches are 
standing and traveling waves \cite{re:golubitsky88}. Examples of the
temporal evolution of eigenvectors of both types are plotted in Fig. 
\ref{ev_conv}. In practice, the specific values of the parameters
determine the way the standing and traveling waves bifurcate
\cite{re:crawford91}. If both 
are supercritical, the one with larger amplitude would be stable, the 
other one unstable.

We finally show our results concerning the location of the oscillatory 
instability boundary as a function of $q$ for two values of $\beta_1$ in 
Fig. \ref{conv}. Stable regions are located to the right of the plotted curves. 
As expected, the stability boundary moves toward the 
Eckhaus line with decreasing $\beta_{1}$, presumably merging with it
for $\beta_{1} \rightarrow 0$. 
Note that for $\beta_{1}=0.5$ the unstable region covers most of the region
of existence of the base states except for a narrow stripe close to the saddle 
node bifurcation.

\section{Discussion}

The coupled system of equations describing fast surface oscillations and
slowly evolving mean flows in three dimensions has been derived. Mean
flows are forced by the surface waves by various
mechanisms. For a particular choice of geometry and limit of
parameters which are relevant to recent experiments, we have shown
that two contributions appear at the appropriate order in the multiple
scale expansion: A viscous streaming flow forced at the free surface
with components of both similar and larger scale compared to the
length scale of the surface
waves, and a long wavelength component originating from slow distortions of
the surface elevation that exists even in the absence of viscosity.

The analysis presented has illustrated the importance of
mean flows in the Faraday wave system for small viscous damping by
determining the stability boundaries of the base pattern of standing
waves against long wavelength perturbations. Since the full system of
surface wave/mean flow equations is quite involved,
we have instead carried out the stability analysis of a phenomenological order
parameter equation, similar in spirit to the Swift-Hohenberg model of
Rayleigh-B\'enard convection. In addition, we have limited the analysis to
the simplest regular pattern consisting of stripes. Mean flows are
induced by perturbation of the stripes, and their coupling to the
order parameter equation affects the stability of stripe solutions. We
have found that the mean flows generally destabilize the base
solution. The strongest coupling, and hence the strongest
destabilization, occurs for longitudinal or Eckhaus
perturbations. Furthermore, mean flows introduce a
new oscillatory instability which for small nonlinear damping in the 
phenomenological model renders all stripe patterns unstable. A weaker 
effect has been found for finite wavelength transverse amplitude
modulations which largely couple only to the short wave part of the
mean flow.

Our first remark concerning experiments follows from the existence of
a longitudinal oscillatory instability. Within the phenomenological
model, all stripe solutions which are stable in the absence of mean
flows are unstable against longitudinal oscillatory perturbations for
sufficiently large coupling parameter $\beta_1$. One would then expect time
dependent behavior at onset. The eigenvectors corresponding to the
oscillatory instability (Fig.~\ref{ev_conv}) show that the
associated mean flow consists of large scale rolls with their axis
oriented parallel to the surface. At the surface it advects the waves
leading to compression and dilation of waves similar to Eckhaus
perturbations, but in the form of traveling or standing waves. 
A numerical solution of the coupled order
parameter mean flow equations shows that the compression not only
leads to a decrease in wave amplitude, but can also result in a
complex cycle including the annihilation of stripes, possibly due to a
different instability triggered by the compression.
We also anticipate novel phenomena arising from mean flows if one
considers the slow dynamics of defects in the wave pattern. Defects as
local perturbations of a regular pattern drive mean
flows which in turn affect defect motion.

\section*{Acknowledgments}
This research was partially supported by the Spanish DGI under Grant
BFM2001-2363, and by the U.S. Department of Energy under contract 
DE-FG05-95ER14566.

\appendix
\section{Order parameter model}

We briefly summarize in this Appendix the derivation of the
phenomenological model given by Eq. (\ref{d40}) of
Sec. \ref{sec:phen_model}, first
introduced in ref. \onlinecite{re:zhang95}. We follow the description
originally introduced by
Zakharov \cite{re:zakharov68}, and later of Crawford, Saffman and Yuen
\cite{re:crawford80} in their study of the nonlinear evolution
of deep water gravity waves in an inviscid, incompressible,
and irrotational fluid. Their analysis can be straightforwardly
extended to a parametrically driven fluid, and linear viscous damping
is added in a phenomenological way. All the variables used in this Appendix
are assumed to be dimensional quantities.

The governing equation for the inviscid fluid is,
\begin{equation}
\left( \bnabla^{2} + \partial_{z}^{2} \right) \phi = 0, \quad\quad -
\infty < z < h({\mathbf x},t),
\end{equation}
with boundary conditions at the free surface $z = h({\mathbf x},t)$,
\begin{equation}
\partial_{t} h + \bnabla h \cdot \bnabla \phi = \partial_{z} \phi,
\end{equation}
\begin{equation}
\partial_{t} \phi + \frac{1}{2} \left( \bnabla \phi \right)^{2} +
\frac{1}{2} \left( \partial_{z} \phi \right)^{2} + \left(g_{0} +
g_{z}(t) \right) h = \frac{\sigma}{\rho} \bnabla \cdot 
\left( \frac{\bnabla h}{
  \sqrt{1+(\bnabla h)^{2}}} \right),
\end{equation}
with $\sigma$ the interfacial tension and
$\rho$ the density of the fluid that is being vibrated with
acceleration $\left( -g_{0}-g_{z}(t) \right)$ in the $z$ direction. 
It is well known that this problem admits a Hamiltonian formulation with
Hamiltonian
\begin{equation}
H = \frac{1}{2} \int d{\mathbf x} \int_{-\infty}^{h({\mathbf x},t)} dz \left[
  \left( \bnabla \phi \right)^{2} + \left( \partial_{z}\phi \right)^{2}
  + \frac{1}{2} \left( g_{0} + g_{z}(t) \right) h^{2} + 
\frac{\sigma}{\rho} \left(
  \sqrt{1+(\bnabla h)^{2}} -1 \right) \right],
\end{equation}
where the velocity potential further satisfies the boundary condition
$\partial_{z} \phi = 0$ as $z \rightarrow - \infty$. The canonically
conjugate variables are the surface deflection $h({\mathbf x},t)$ and the
velocity potential on the surface $\phi^{s}({\mathbf x},t) =
\phi({\mathbf x}, z=h({\mathbf x}))$.
Phenomenological damping can be introduced by considering a
dissipation function $Q(h({\mathbf x},t),\phi^{s}({\mathbf
  x},t))$. The resulting canonical equations of motion are,
\begin{equation}
\partial_{t} h({\mathbf x},t) = \frac{ \delta H}{\delta
  \phi^{s}({\mathbf x},t)},
\label{eq:hamilton1}
\end{equation}
\begin{equation}
\partial_{t} \phi^{s}({\mathbf x},t) = - \frac{\delta H}{\delta
  h({\mathbf x},t)} + Q(h({\mathbf x},t),\phi^{s}({\mathbf x},t)).
\label{eq:hamilton2}
\end{equation}
The functional $Q$ determines the rate of viscous
dissipation in Eqs. (\ref{eq:hamilton1})-(\ref{eq:hamilton2}) so that
\begin{equation}
\frac{dH}{dt} - \frac{\partial H}{\partial t} = \int d{\mathbf x}
Q(h({\mathbf x},t),\phi^{s}({\mathbf x},t)) \partial_{t} h({\mathbf x},t).
\label{eq:dissipation}
\end{equation}
Of course, $Q=0$ corresponds to the inviscid limit.

The case of a fluid of low viscosity has been treated by assuming that
energy dissipation is dominated by potential flow in the bulk
\cite{re:landau76}. The functional $Q$ can then be
determined by equating the rate of dissipation in
Eq. (\ref{eq:dissipation}) to the rate of energy dissipation due to potential flow,
\begin{equation}
\int d{\mathbf x}
Q(h({\mathbf x},t),\phi^{s}({\mathbf x},t)) \partial_{t} h({\mathbf x},t) = - \nu
\int d{\mathbf x} \int_{\infty}^{h({\mathbf x},t)} dz \nabla^{2} (\nabla
\phi)^{2}.
\end{equation}
This equation has been used to determine $Q$ order by order in an expansion
in the surface wave steepness \cite{re:milner91,re:lyngshansen97}. To
the order relevant here, one finds,
\begin{equation}
\hat{Q}({\mathbf k},t) = - 4 \nu k^{2} \hat{\phi}^{s}({\mathbf k},t) + {\rm
  nonlinear ~ terms},
\end{equation}
where $\hat{Q}({\mathbf k},t)$ is the Fourier transform of $Q$.
As discussed in refs. \onlinecite{re:zhang94,re:zhang97}, this
approximation yields, by construction, the correct rate of energy
dissipation at linear order, but not the correct equations of motion
even at this order. In particular, it overestimates by a factor of two
the damping force in Eq. (\ref{eq:hamilton2}), and omits wave
rectification in Eq. (\ref{eq:hamilton1}) that arises from the rotational
component of the flow in a thin boundary layer adjacent to the free surface.

Following Zakharov \cite{re:zakharov68}, we define a complex field
\begin{equation}
b({\mathbf k},t) = \sqrt{\frac{\omega(k)}{2k}} \hat{h}({\mathbf k},t)
+ i \sqrt{\frac{k}{2 \omega(k)}} \hat{\phi}^{s}({\mathbf k},t) ,
\end{equation}
where $\hat{h}({\mathbf k},t)$ and $\hat{\phi}^{s}({\mathbf k},t)$ are the two
dimensional Fourier transforms of $h({\mathbf x},t)$ and
$\phi^{s}({\mathbf x},t)$ respectively, and $\omega(k) = \sqrt{g_{0}k+
\sigma k^{3}/\rho}$ is the inviscid dispersion relation. In terms of this
new variable, the Hamiltonian system
(\ref{eq:hamilton1})-(\ref{eq:hamilton2}) can be written as,
\begin{equation}
\partial_{t} b({\mathbf k},t) = - i  \frac{\delta H}{\delta
  \bar b(-{\mathbf k},t)} + i \sqrt{
  \frac{k}{2 \omega(k)}} \hat{Q}({\mathbf k},t).
\label{eq:b_eq}
\end{equation}
Equation (\ref{eq:b_eq}) is now expanded in a power series of $b$. We
confine ourselves here to linear terms in $b$, as nonlinear
terms will be added phenomenologically. However, explicit forms of
cubic terms in $b$ have been obtained \cite{re:zhang94b} both for the
present case of an expansion around the inviscid solution, and also
for the Linear Damping Quasi Potential Equations of ref. 
\onlinecite{re:zhang97}.

By expanding Eq. (\ref{eq:b_eq}) in power series of $b$, we find,
\begin{equation}
\partial_{t} b({\mathbf k},t) + 2 \nu k^{2} \left[ 
b({\mathbf k},t) - \bar{b}(-{\mathbf k},t) \right] + i \omega(k)
b({\mathbf k},t) + \frac{i k g_{z}(t)}{2 \omega(k)} \left[ b({\mathbf
  k},t) + \bar{b}(-{\mathbf k},t) \right] + {\cal NL}(b({\mathbf k},t)) =
0,
\label{eq:bqt}
\end{equation}
where ${\cal NL}(b({\mathbf k},t))$ stands for terms nonlinear in the
amplitudes $b$.

If the driving acceleration is given by $g_{z}(t) = a \cos \Omega t$,
only amplitudes with wavenumber close to the critical
wavenumber $k_{0}$ are excited near onset, with frequency close to the resonant
frequency $\omega(k_{0})=\Omega/2$. We introduce a conventional
multiple scale expansion near onset, but choose to do so so in a
manner that will preserve the rotational invariance of the original
governing equations. We further assume the following scalings for the
damping and driving terms: $\gamma^{\prime}= 2 \nu k_{0}^{2} = 
\epsilon^{2} \gamma_{0}$
and $f^{\prime} = k_{0} a/ (4 \omega(k_{0})) = \epsilon^{2} f_{0}$, where
$\epsilon$ is a small expansion parameter, and both $\gamma_{0}$ and
$f_{0}$ are ${\cal O}(1)$ quantities. We also expand
\begin{equation}
b({\mathbf k},t) = \epsilon B({\mathbf k},T_{1},T_{2}) e^{-i
 \omega(k_{0})t} + \epsilon^{2} b_{2}({\mathbf k},t)+ \epsilon^{3} 
b_{3}({\mathbf k},t) + \ldots
\label{eq:appex}
\end{equation}
with $T_{1} = \epsilon t$ and $T_{2} = \epsilon^{2} t$. The slow time
scale $T_{1}$ corresponds to the time scale of translation of a wave
packet, whereas $T_{2}$ is the scale of change in the modulation of
the wave packet. These two time scales are consistent with an
expansion of the inviscid dispersion relation $\omega(k) =
\omega(k_{0}) + \epsilon \omega^{\prime} +
\epsilon^{2} \omega^{\prime\prime} + \ldots$ for modes near the
critical wavenumber $k_{0}$. Substitution of Eq. (\ref{eq:appex}) into
Eq. (\ref{eq:bqt}) shows that the equation is identically satisfied at
${\cal O}(\epsilon)$. At ${\cal O}(\epsilon^{2})$ we obtain the
following solvability condition,
\begin{equation}
\frac{\partial B}{\partial T_{1}} = - i \omega^{\prime} B.
\end{equation}

The solvability condition at order ${\cal O}(\epsilon^{3})$ is
\begin{equation}
\frac{\partial B}{\partial T_{2}} = - \gamma_{0} B({\mathbf k}) - i
f_{0}  \bar B(-{\mathbf k}) - i \omega^{\prime\prime}  B({\mathbf k}) +
{\cal NL}[B].
\end{equation}
with a known nonlinear functional $ {\cal NL}[B]$. We now combine the two
solvability conditions by writing $A({\mathbf k}) = \epsilon B ({\mathbf
k} )$ and $\partial_{t} A = \epsilon^{2} \partial_{T_{1}} B +
  \epsilon^{3} \partial_{T_{2}} B$, and find,
\begin{equation}
\frac{\partial A}{\partial t} = - \gamma^{\prime} A({\mathbf k},t) - i
f^{\prime}  \bar A(-{\mathbf k},t) - i \left(\epsilon \omega^{\prime} + 
\epsilon^{2} \omega^{\prime\prime}\right) A({\mathbf k},t) + {\cal NL}[A]
\label{eq:aqt}
\end{equation}
Hence the slow evolution near onset given by Eq. (\ref{eq:aqt}) is the same as
that of the original set of inviscid equations supplemented by phenomenological
linear damping. 

From the inviscid dispersion relation, we find,
\begin{equation}
\epsilon \omega^{\prime} + \epsilon^{2} \omega^{\prime\prime} = c_{1} 
\left( k^{2} -
k_{0}^{2} \right) + c_{2} \left( k^{2} - k_{0}^{2} \right)^{2} + {\cal
  O}((k-k_{0})^{3}),
\end{equation}
with
\begin{equation}
c_{1} = \frac{g_{0}}{4k_{0}\omega(k_{0})} + \frac{3 \sigma k_{0}}{4 \rho
  \omega(k_{0})},
\end{equation}
and
\begin{equation}
c_{2} = \frac{1}{4 k_{0}^{2} \omega(k_{0})} \left( \frac{3 \sigma
  k_{0}}{4\rho} - \frac{(g_{0}+ 3 \sigma k_{0}^{2}/\rho)^{2}}{8
  \omega(k_{0})^{2}} - \frac{g_{0}}{4k_{0}} \right).
\end{equation}

We derive next a real space order parameter model from
Eq. (\ref{eq:aqt}). Three simplifications are necessary. First,
the nonlinear functional in Eq. (\ref{eq:aqt}) does not have a closed
form representation in real space. As has been done in other systems
(cf. Rayleigh-B\'enard convection, \cite{re:swift77}), we introduce
phenomenological functional forms for this term. In doing so we
artificially determine the symmetry of the bifurcating pattern at
onset, but more importantly in the case of Faraday waves, we sidestep the
issue of the origin of nonlinear damping and saturation of the
waves \cite{re:zhang97,re:chen97}. In the simplest possible case,
the nonlinear term in Eq. (\ref{eq:aqt}) is approximated by an imaginary
constant $iR^{\prime}$. Second, it is also known that linear damping is not
sufficient to produce wave saturation in this system
\cite{re:milner91}. We introduce a phenomenological nonlinear damping
coefficient $\alpha \gamma^{\prime}$, where $\alpha$ is a constant
assumed to be of order 1.

We finally define a complex order parameter field $\psi({\mathbf x},t)$
as the inverse Fourier transform of $A({\mathbf k})$, and find from
Eq. (\ref{eq:aqt}),
\begin{equation}
\partial_{t} \psi = - \gamma^{\prime} \psi + i f^{\prime} \bar\psi + i c_{1} \left(
k_{0}^{2} + \bnabla^{2} \right) \psi - i c_{2} \left( \bnabla^{2} +
k_{0}^{2} \right)^{2} \psi + (-\alpha \gamma^{\prime} + i R^{\prime}) 
\| \psi \|^{2} \psi .
\end{equation}
We now choose $\omega(k_{0})=\Omega/2$ as the unit of time,
$1/k_{0}$ as the unit of length, and further define $f =
f^{\prime}/\omega(k_{0}) = k_{0}a/4\omega(k_{0})^{2}$,
$\gamma = \gamma^{\prime}/\omega(k_{0}) = 2 \nu k_{0}^{2}/\omega(k_{0})$,
and $R = R'/\omega(k_{0})$. By choosing $1/\sqrt{R}$ as the order parameter scale,
Eq. (\ref{d40}) results in the capillary wave limit.
The positive sign of the imaginary part of the nonlinear coefficient
$(- \alpha \gamma + i)$ in Eq. (\ref{d40}) is chosen to represent
capillary waves. In the opposite limit of gravity waves, the imaginary
part of this coefficient has to be negative. Note that as a
third simplification we have eliminated the term
$i\left(1+\bnabla^{2}\right)^{2} \psi$ in Eq. (\ref{d40}) as this term
together with $i
\left( 1 + \bnabla^{2} \right) \psi$ leads to two different wavenumbers
becoming critical at threshold, and unwanted feature for us. 

The effect of this third simplification can be further understood by
comparing the amplitude equation on the model and that of the
inviscid fluid. By introducing a multiple scale expansion of the form
\begin{equation}
\psi = \delta \sum_{j=\pm 1}^{\pm N} a_{j}({\mathbf X},t,T_{1},T_{2}) 
e^{i \hat{\mathbf k}_{j}
  \cdot {\mathbf x}} + \delta^{2} \psi_{2} + \delta^{3} \psi_{3} \ldots
\end{equation}
with $\delta$ a small bookkeeping parameter, and ${\mathbf X} = \delta
{\mathbf x}$, $T_{1} = \delta t$, and $T_{2} = \delta^{2} t$ we find
that up to order ${\cal O}(\delta^{3})$ \cite{re:zhang94}
$$
\partial_{t} a_{j} = - \gamma a_{j} + i f \bar a_{-j} - \frac{3}{2}
\left( \hat{\mathbf k}_{j} \cdot \bnabla \right) a_{j} + \frac{3i}{4}
\bnabla^{2} a_{j} + 
$$
\begin{equation}
+ \left( - \alpha \gamma + i \right) \left(
|a_{j}|^{2} a_{j} + 2 \sum_{l\neq j}|a_{l}|^{2} a_{j} + 2 \sum_{l\neq
    \pm j} a_{l} a_{-l}\bar a_{-j} \right)
\label{eq:ampli_eq_model}
\end{equation}
by following the same expansion procedure outlined above. The terms
linear in the amplitudes are the same as the corresponding terms in
the amplitude equation derived directly from the inviscid equations
except for an additional term 
$(\hat{\mathbf k}_{j}\cdot\bnabla)^{2} a_{j}$ which is missing
in Eq. (\ref{eq:ampli_eq_model}). This is a direct consequence of having
eliminated the term $i\left(1+\bnabla^{2}\right)^{2} \psi$ in the 
phenomenological model.

\bibliographystyle{apsrev}
\bibliography{$HOME/mss/references}

\begin{figure}[t]
\caption{This figure illustrates the weak destabilization of the base
  solution against Transverse Amplitude Modulations due to the
  coupling to mean flows. Shown are the critical values for
  instability $\mu_{crit}$ arising from either the long
  wavelength component of the mean flow velocity alone (\lq\lq ms'') or the
  short wavelength component alone (\lq\lq mo'') as a function of the
  coupling parameter $\beta_{1}$. The base solution is stable above
  the corresponding lines. The figure shows that the \lq\lq ms''
  component of the flow does not appreciably modify the stability
  threshold, whereas the effect of the \lq\lq mo" component is to
  weakly destabilize the base solution with increasing $\beta_{1}$. The
  wavenumber of the base solution is $q=1.04$, and we have used
  $\gamma = 0.1, \alpha=0.5$, and $\Gamma = 0.8$. The parameter
  $\beta_{1}$ spans the range between no mean flow and the approximate
  value that corresponds to the weak viscosity but shallow layer
  experiments of ref. \protect\cite{re:kudrolli96}. The values of
  $\beta_2$ and $\lambda$ depend on $\beta_1$ according to Eq. (\ref{beta}).}
\label{comp}
\end{figure}

\begin{figure}[t]
\caption{Stability diagram for (a) the order parameter model of 
Eq.~(\ref{d40}) without mean flow, and (b) with $\beta_1=0.5$. Other
values of the parameters used are $\gamma=0.1, \alpha=0.5,$ and 
$\Gamma=0.8$. 
We show the Eckhaus line (dotted), TAM line (dashed), and the zig-zag line
(dash-dotted). We also show the neutral stability curve of the primary instability
 (solid line denoted
by N), and a saddle node bifurcation (thick solid line marked S).
Only the left branch of the Eckhaus line is shown emanating from 
$(q=1,\mu=0)$. This line is parabolic near the critical point,
but quickly bends to the right as shown in the figure. Hence the region
of stability of the base solution against an Eckhaus instability is the
region below the dotted line. Comparison of (a) and (b) shows that the mean
flow decreases the regions of stability against Eckhaus modulations
and to a small extent against transverse amplitude modulations. 
In (a) the region of stability against all perturbations
is shown by the gray area. This region is not indicated in (b) since these
solutions are unstable with respect to the oscillatory instability.}
\label{balloon}
\end{figure}

\begin{figure}[t]
\caption{Real (left) and imaginary (right) part of the largest eigenvalue 
for $\beta_1=0.2$, $\mu=0.1318$ and $q=1.07$. The inset shows the
region near $k=0$.}
\label{ew}
\end{figure}

\begin{figure}[t]
\caption{Critical eigenvectors corresponding to
the oscillatory instability at $\beta_1=0.2$, $q=1.0$,
$\mu=0.1318$, and $k=0.04$.
We show the temporal evolution  over its own period (vertical axis)
to illustrate the nature of the critical modes.
The eigenvectors are represented by  the $x$ component of ${\bf u}^{ms}$ (a,c)
and the amplitude of $\psi$ (b,d). The evolution is obtained 
from the two critical eigenvectors $v_{i}$ of the complex matrix $A(q,k,\dots)$  with eigenvalues $\pm i \omega$. Left and right traveling (a,b) waves are given 
by the evolution of $v_1 e^{i \omega t}$ and $v_2 e^{-i \omega t}$, respectively.
A standing wave is obtained (c,d)
by superposition of the evolution from both eigenvectors scaled to equal amplitude. The actual solution for $\psi$ is obtained from the superposition of
the plotted eigenvectors with the base solution, see Eq.~(\ref{eq:psi_pert}). }
\label{ev_conv}
\end{figure}

\begin{figure}[t]
\caption{Stability boundaries of the oscillatory instability for three values 
of $\beta_1$: 0.05 (dotted line), 0.2 (dashed line), and 0.5 (solid line). 
The thick solid line indicates the saddle node. Periodic solutions are unstable
to the left of the curves. Comparison with Fig.~\ref{balloon} shows, that all
periodic solutions are unstable and the pattern is expected to be time-dependent.}
\label{conv}
\end{figure}

\end{document}